\documentclass[prb,twocolumn,showpacs,preprintnumbers,amsmath,amssymb,superscriptaddress]{revtex4}

\usepackage{graphicx}
\usepackage{dcolumn}
\usepackage{bm}
\def\Vec#1{\bm{#1}}

\def\up{\uparrow}
\def\down{\downarrow }

\begin{document}


\title{
Impurity Scattering Rate and Coherence factor in Vortex Core of Sign-reversing 
$s$-wave Superconductors
}

\author{Yuki Nagai}
\affiliation{CCSE, Japan Atomic Energy Agency, 6-9-3 Higashi-Ueno, Tokyo 110-0015, Japan}
\affiliation{
Department of Physics, The University of Tokyo, Tokyo 113-0033, Japan
}
\affiliation{CREST(JST), 4-1-8 Honcho, Kawaguchi, Saitama, 332-0012, Japan}
\affiliation{JST, TRIP, Chiyoda, Tokyo, 102-0075, Japan}

\author{Yusuke Kato}
\affiliation{
Department of Basic Science, The University of Tokyo, Tokyo 153-8902, Japan
}

\date{\today}

\begin{abstract}
We investigate the impurity scattering rates for quasi-particles in vortex cores of sign-reversing $s$-wave superconductors 
as a probe to detect the internal phase difference of the order parameters among different Fermi surfaces. 
The impurity scattering rates and coherence factors are related to quasiparticle interference effect by the scanning tunneling microscopy and spectroscopy technique. 
With use of the 
Born and Kramer-Pesch approximations for the Andreev bound states, 
we 
show that 
the sign-reversed forward scatterings are dominant in vortex cores. 
Owing to the coherence factor in vortex cores of $\pm s$-wave superconductors, the impurity scattering rate of 
the Andreev bound states 
has a characteristic distribution 
on the Fermi surfaces. For comparison, the impurity scattering rates in vortex cores of $s$-wave and $d$-wave superconductors are also discussed. 
\end{abstract}

\maketitle
\section{Introduction}
The discovery of novel Fe-based superconductors has attracted 
considerable attention because of high superconducting transition temperature.\cite{Kamihara} 
A $\pm s$-wave pairing symmetry has been theoretically proposed as a candidate for the pairing symmetry in the Fe-based superconductors.
\cite{bang0807,parish,kuroki,seo,Arita,mazin,eremin,nomura,stanev,Senga} 
What we mean by superconductors with the $\pm s$ wave pairing symmetry is multi-band superconductors the pair-potentials of which are $s$-wave within a Fermi surface but change in sign between different Fermi surfaces. 
A detection of a sign change in the 
pair potential between the Fermi surfaces would provide a definitive evidence for the $\pm s$-wave pairing-symmetry. It is, however, difficult to detect such a relative phase in spatially uniform systems. In order to develop a ^^ ^^ phase-sensitive" probe, the introduction of the inhomogeneities such as interface, vortices and impurities are crucial. 
%

As a phenomenon sensitive to the relative phase, 
the formation of the Andreev bound states at a surface or junction 
has been discussed in many papers\cite{Hu,TanakaPRL,Kashiwaya,Choi, Golubov, Araujo,Wang,Linder,LinderPRB,Tsai,Ghaemi,Zare}. With use of this phenomenon, a phase-sensitive experiment has been proposed in Refs.~\onlinecite{NagaiPRB09, OnariTanaka} to detect the $\pm$ $s$-wave pairing-symmetry in the Fe-based superconductors.
%
%

Another phase-sensitive experiment, which is related to scattering off impurities and vortices, is the measurement of quasiparticle interference (QPI) patterns by the scanning tunneling microscopy/spectroscopy (STM/STS) in superconductors\cite{Hoffman,McElroy, HanaguriNphys,Kohsaka,Hanaguri,Pereg-Barnea08,Maltseva}. 
The QPI was originally meant for the standing wave coming from elastic scattering of electrons by steps or point disorders on the surfaces of metals or semiconductors and was firstly observed on the surfaces of Au\cite{Hasegawa93} and Cu\cite{Crommie93}. Subsequently, the QPI was utilized to deduce the quasiparticle spectrum of a cuprate superconductor Bi$_2$Sr$_2$CaCu$_2$O$_{8+\delta}$ with $d$-wave pairing symmetry. 
Furthermore, Hanaguri {\it et al. } have measured the dependence on magnetic fields in QPI patterns in another cuprate superconductor Ca$_{2-x}$Na$_{x}$CuO$_{2}$Cl$_{2}$\cite{Hanaguri} and successfully worked out the coherence factor of each quasi-particle scattering from the spatial Fourier-transform of the ratio $Z(\Vec{r},V)=g(\Vec{r},V)/g(\Vec{r},-V)$ of the conduction maps $g(\Vec{r},V)=dI/dV(\Vec{r},V)$ for the position $\Vec{r}$ and the bias voltage $\pm V$. The field dependence of QPI in $d$-wave superconductors has been theoretically studied in Refs.~\onlinecite{Maltseva,Pereg-Barnea08}. 
The coherence factor is a factor appearing in a scattering matrix element of quasi-particles in superconductors and it is sensitive to the momentum dependence of the phase of the pair-potential and the type of scatterers (non-magnetic scatterers, magnetic scatterers or the spatial variation of the pair-potentials). 
The QPI measurement is now becoming a brand-new phase-sensitive tool. 

Recently, the QPI measurements in magnetic fields have been done for the Fe-based superconductors Fe(Se,Te) \cite{Hanaguri2010}. Their results suggest the $\pm s$-wave pairing symmetry in those materials. The QPI for the Fe-based superconductors has been discussed theoretically in Ref.~\onlinecite{Zhang}. 
%
%
%
%
%
%

In $\pm s$-wave superconductors, the energy spectrum is fully-gapped in the absence of magnetic fields
and 
the low energy excitations in the presence of magnetic fields are exhausted by the localized modes around vortex cores (the Caroli-deGennes-Matricon mode\cite{Caroli64}). Those localized modes can be regarded as the Andreev bound states when the quasiclassical condition (that the coherence length is much larger than the Fermi wave-length) is satisfied\cite{Kramer74,Rainer96,Stone96,KopninText}. For the $\pm$ $s$-wave superconductors, so far, (i) the contribution to the QPI from those localized modes has not been considered and (ii) neither the impurity scattering rate nor the coherence factor of the localized modes around vortex cores has been derived explicitly. (iii) It has not been studied whether the impurity scattering has nontrivial aspects near vortex cores of $\pm s$ wave superconductors. The latter two issues (ii) and (iii) are addressed in the present paper.   


The aim of this paper is two-fold. First we derive explicit expressions for the impurity scattering rate and the related coherence factor of the Andreev bound states localized near vortex cores. Second, we present a novel scattering property inherent to the Andreev bound states in $\pm$ $s$-wave superconductors. 

As a main result, we will show that the coherence factor of impurity scattering inside vortex cores is different from that for quasi-particles in the bulk. The impurity scattering rate $\Gamma_{\Vec{k},\Vec{k'}}$ from the initial state $\Vec{k}$ to the final state $\Vec{k}'$ is large when the following conditions are satisfied. (i) The signs of the pair-potentials for $\Vec{k}$ and $\Vec{k}'$ are different and (ii) the Fermi velocities $\Vec{v}(\Vec{k})$ and $\Vec{v}(\Vec{k}')$ is similar. We call the scattering for which (i) and (ii) are satisfied ^^ ^^ the sign-reversal forward scattering". As another result, we will demonstrate that those sign-reversal forward scatterings can occur for quasi-particles in a substantial portion of the Fermi surfaces near vortex cores in the $\pm s$-wave superconductors. 

The formulation and approximation which we will use are outlined below. The impurity scattering rates of the Andreev bound states in vortex cores have been treated analytically \cite{Larkin76,Kopnin99,Kato,Kato03} within the quasiclassical Eilenberger theory\cite{Eilenberger68,Larkin68} for $s$-wave\cite{Larkin76,Kopnin99} and chiral $p$-wave superconductors.\cite{Kato,Kato03}  
Those analytical results on the impurity scattering rates have been confirmed to be consistent with numerical results of self-consistent Born approximation.\cite{Kato02,Tanuma09}  
The analytical calculation of the Andreev bound states in vortex cores originates from the paper by Kramer and Pesch \cite{Kramer74} in pure superconductors and has been generalized to the impure case.\cite{Larkin76,Kopnin99,Kato} We thus call the method used in Refs.~\onlinecite{Kopnin99,Kato} the Kramer-Pesch approximation in this paper. 
The quasiclassical theory is applicable to Fe-based superconductors since 
the band widths ($\sim$ a few eV) are much larger than the maximum of the superconducting gap ($\sim 10$meV). 
The iron-based superconductors are known to be multi-band systems and have multiple Fermi surfaces.\cite{kuroki} %
Assuming that the intraband pairings are dominant, one can define the quasiclassical Green functions on each band. 
One can then solve the equation of motion of the quasiclassical Green function  near a vortex core and 
can obtain the analytical expression for the impurity scattering rate with the 
Born approximation in multi-band systems by extending the method developed in Refs.~\onlinecite{Kopnin99,Kato}
%
%
%

This paper is organized as follows. 
The Born approximation and the quasiclassical approximation for multi-band superconductors are presented in Sec.~II. 
The generalization to multi-band superconductors is shown to be straightforward
when we assume that the intraband pairings are dominant. 
The impurity self-energy in a single vortex within the Kramer-Pesch approximation is derived in Sec.~III. 
The analytical expression for the impurity scattering rates $\Gamma_{\Vec{k},\Vec{k'}}$ in vortex cores is derived. 
The results are shown in Sec.~IV. 
We discuss the coherence effects on the scattering-angle dependence for the cases of the sign-conserved and the sign-reversed scatterings. 
We also calculate the $\Vec{q} \equiv \Vec{k}' - \Vec{k}$-dependence of the impurity scattering rate for the isotropic $s$-wave, $d$-wave and 
the isotropic $\pm s$-wave superconductors, respectively. 
The conclusion is given in Sec.~V. 
\section{Born Approximation and Quasiclassical Approach}
\subsection{Orbital representation and Band representation}
Let us consider two-dimensional superconductors. 
We consider an $n$-orbital system which is a periodic crystal with $n$ atomic orbitals in a unit cell. 
We introduce a Hamiltonian written as
\begin{equation}
H = \sum_{\Vec{k},\sigma,\mu,\nu} \epsilon_{\Vec{k},\mu,\nu} c_{\Vec{k} \mu \sigma}^{\dagger}c_{\Vec{k} \nu \sigma} 
+ \sum_{\Vec{k},\mu,\nu} \left( \Delta_{\Vec{k},\mu,\nu} c_{\Vec{k} \mu \up}^{\dagger} c_{-\Vec{k} \nu \down}^{\dagger} + {\rm h.c.} \right),
\end{equation}
where the operator $c_{k \mu \sigma}^{\dagger}$ creates an electron with spin $\sigma$ and momentum $\Vec{
k}$ 
on the $\mu$-th orbital. 
In the matrix form, this Hamiltonian is a $2n \times 2n$ matrix in Nambu and orbital spaces 
expressed as 
\begin{eqnarray}
\check{H}^{\rm o}(k_{x},k_{y}) =
\left(\begin{array}{cc}
\hat{H}^{\rm o}(k_{x},k_{y}) & \hat{\Delta}^{\rm o}(k_{x},k_{y})  \\
\hat{\Delta}^{\rm o \dagger}(k_{x},k_{y})  & - \hat{H}^{\rm o}(k_{x},k_{y}) 
\end{array}\right),
\end{eqnarray}
in the ``orbital representation'' where the basis functions are atomic orbitals in a unit cell. 
Here $\hat{H}^{\rm o}$ is the Hamiltonian in the normal state represented as an $n \times n$ 
matrix in the orbital space and 
$\hat{\Delta}^{\rm o}$ is the superconducting order parameter. 
From now on, the subscript ``o'' indicates that matrices are represented with the orbital basis, 
$hat$ $\hat{a}$ denotes an $n \times n$ matrix in the orbital space and $check$ $\check{a}$ denotes 
a $2 n \times 2 n$ matrix composed of the $2 \times 2$ Nambu space and the $n \times n$ orbital space. 
The unperturbed $2 n \times 2 n$ Green function in the orbital representation is defined as 
\begin{equation}
\check{G}_{0}^{\rm o}(k_{x},k_{y}; i \omega_{n}) = [i \omega_{n} \check{1} - \check{H}^{\rm o}(k_{x},k_{y})]^{-1}. \label{eq:bulk}
\end{equation}
Here $\omega_{n}$ is the Fermion Matsubara frequency. 

We also introduce an $n \times n$ Hamiltonian in the ``band representation'' defined by 
\begin{eqnarray}
\hat{H}^{\rm b}(k_{x},k_{y}) &\equiv& \hat{P}^{-1}(k_{x},k_{y}) \hat{H}^{\rm o}(k_{x},k_{y})\hat{P}(k_{x},k_{y}), \\
 &=& 
 \left(\begin{array}{ccc} \lambda_{1} & 0 & 0\\ 
 0 & \ddots & 0\\
0 &0 &\lambda_{n} \end{array}\right).
\end{eqnarray}
Here $\lambda_{i}$ ($i=1,2,\cdots,n$) denotes the $i$-th largest eigenvalue. %
$\hat{P}$ is a unitary matrix consisting of the eigenvectors for the Hamiltonian $\hat{H}^{\rm o}(k_{x},k_{y})$. 
The $2n \times 2n$ Hamiltonian in Nambu and orbital spaces in the ``band representation''  is also defined by 
\begin{eqnarray}
\check{H}^{\rm b}(k_{x},k_{y}) &\equiv& 
\check{U}^{-1}(k_{x},k_{y}) \check{H}^{\rm o}(k_{x},k_{y}) \check{U}(k_{x},k_{y}), \label{eq:bhami}\\
&=&  \left(\begin{array}{cc}\hat{H}^{\rm b} & \hat{\Delta}^{\rm b} \\ \hat{\Delta}^{{\rm b} \dagger} & - \hat{H}^{\rm b} \end{array}\right),
\end{eqnarray}
where 
\begin{eqnarray}
\check{U}(k_{x},k_{y}) &\equiv&  
\left(\begin{array}{cc}\hat{P}(k_{x},k_{y}) &0 \\ 
0& \hat{P}(k_{x},k_{y}) \end{array}\right), \\
\hat{\Delta}^{\rm b} &\equiv& \hat{P}^{-1} \hat{\Delta}^{\rm o} \hat{P}.
\end{eqnarray}
The unperturbed $2 n \times 2 n$ Green function in the band representation is written as 
\begin{equation}
\check{G}_{0}^{\rm b}(k_{x},k_{y};i \omega_{n}) = \check{U}(k_{x},k_{y})^{-1} \check{G}_{0}^{\rm o}(k_{x},k_{y};i \omega_{n}) 
\check{U}(k_{x},k_{y}). \label{eq:greenb}
\end{equation}
In general, $\hat{\Delta}^{\rm b}$ contains off-diagonal elements, which correspond to inter-band pairings.

\subsection{Born approximation}
We consider the Green function under the influence of a lot of weak impurities. 
The Green function with the Born approximation in the orbital representation is 
written as 
\begin{eqnarray}
\check{G}^{\rm o}(\Vec{k}) = \check{G}^{\rm o}_{0}(\Vec{k})  + 
\check{G}^{\rm o}_{0}(\Vec{k}) \check{\Sigma}^{\rm o}(\Vec{k})  \check{G}^{\rm o}_{0}(\Vec{k}), \label{eq:borngreeno}
\end{eqnarray}
where 
\begin{eqnarray}
\check{\Sigma}^{\rm o}(\Vec{k}) &\equiv& n_{\rm imp} \int  \frac{d \Vec{k}_{1} }{(2 \pi)^{2}} \check{u}(\Vec{k}-\Vec{k}_{1}) 
\check{G}^{\rm o}_{0}(\Vec{k}_{1})\check{u}(\Vec{k}_{1}-\Vec{k}), \: \: \: \: \: \\
\check{u}(\Vec{k})&\equiv& \left(
\begin{array}{cc}
 \hat{u}(\Vec{k}) &  0 \\
0 &- \hat{u}(\Vec{k})
\end{array}
\right).
\end{eqnarray}
Here $n_{\rm imp}$ is the concentration of impurity atoms, $\hat{u}(\Vec{k})$ is the impurity potential.  
By substituting eq.~(\ref{eq:greenb}) into eq.~(\ref{eq:borngreeno}), the Green function in the band representation is written as 
\begin{eqnarray}
\check{G}^{\rm b}(\Vec{k}) = \check{G}^{\rm b}_{0}(\Vec{k})  + 
\check{G}^{\rm b}_{0}(\Vec{k}) \check{\Sigma}^{\rm b}(\Vec{k})  \check{G}^{\rm b}_{0}(\Vec{k}), \label{eq:borngreenb}
\end{eqnarray}
where 
\begin{eqnarray}
\check{\Sigma}^{\rm b}(\Vec{k}) &\equiv& n_{\rm imp} \int  \frac{d \Vec{k}_{1} }{(2 \pi)^{2}} \check{v}(\Vec{k},\Vec{k}_{1}) 
\check{G}^{\rm b}_{0}(\Vec{k}_{1})\check{v}(\Vec{k}_{1},\Vec{k}), \: \: \: \\
\check{v}(\Vec{k},\Vec{k}_{1}) &\equiv&  \check{U}(\Vec{k})^{-1} \check{u}(\Vec{k} - \Vec{k}_{1})
\check{U}(\Vec{k}_{1}). 
\end{eqnarray}
It should be noted that the impurity potential in the band representation $\check{v}(\Vec{k},\Vec{k}_{1})$ is 
a function of $\Vec{k}$ and $\Vec{k}_{1}$, not a function of $\Vec{k} - \Vec{k}_{1}$. 
Assuming that intraband pairings are dominant, we neglect the off-diagonal (interband) elements in $\hat{\Delta}^{\rm b}$. 
In this case, the normal and anomalous parts of the self-energy $\hat{\Sigma}^{{\rm b},{\rm N}}(\Vec{k})$, $\hat{\Sigma}^{\rm b,A}(\Vec{k})$ are expressed as 
\begin{widetext}
\begin{eqnarray}
\hat{\Sigma}^{\rm b,N}_{ij}(\Vec{k}) &=& n_{\rm imp} \int \frac{d \Vec{k}_{1}}{(2 \pi)^{2}} 
\sum_{m} \hat{v}_{im}(\Vec{k},\Vec{k}_{1}) \hat{G}_{0,mm}^{\rm b}(\Vec{k}_{1})\hat{v}_{mj}(\Vec{k}_{1},\Vec{k}), \\
\hat{\Sigma}_{ij}^{\rm b, A}(\Vec{k})
&=& n_{\rm imp} \int \frac{d \Vec{k}_{1}}{(2 \pi)^{2}} 
\sum_{m} \hat{v}_{im}(\Vec{k},\Vec{k}_{1}) \hat{F}_{0,mm}^{\rm b}(\Vec{k}_{1})\hat{v}_{mj}(\Vec{k}_{1},\Vec{k}), 
\end{eqnarray}
\end{widetext}
where\cite{note}, 
\begin{eqnarray}
\hat{v}(\Vec{k},\Vec{k}_{1}) &\equiv& \hat{P}(\Vec{k})^{-1} \hat{u}(\Vec{k}-\Vec{k}_{1}) \hat{P}(\Vec{k}_{1}), \\
\check{G}^{\rm b}_{0} &\equiv& 
- \left(
\begin{array}{cc}
 \hat{G}_{0}^{\rm b} &  \hat{F}_{0}^{\rm b} \\
 \hat{F}_{0}^{\rm b \dagger} &- \hat{\bar{G}}_{0}^{\rm b}
\end{array}
\right).
\end{eqnarray}
Here the symbols $i$,$j$,$m$ denote the band indices. 
For simplicity, we assume that the perturbed Green function $\check{G}^{\rm b}$ is diagonal in the band space. 
For example, the Green function in a two-band system is written as \cite{Senga,Ichioka,Gumann,Graser}
\begin{eqnarray}
\check{G} = \left(\begin{array}{cccc}
G_{\alpha} & 0 & F_{\alpha} & 0 \\
0 & G_{\beta} & 0 & F_{\beta} \\
-F_{\alpha}^{\dagger} & 0 & \bar{G}_{\alpha} & 0 \\
0 & - F_{\beta}^{\dagger} & 0 & \bar{G}_{\beta}
\end{array}\right).
\end{eqnarray}
Under this assumption, 
one can regard the self-energy as diagonal with respect to the band index.

\subsection{Quasiclassical Green functions}
We assume $|\Delta_{ii}^{\rm b}| \ll E_{F}$. 
This relation is satisfied in most of systems such as conventional superconductors and the Fe-based ones 
since the band width is a few eV and the superconducting order parameter is the order of 10meV in these materials. 
In this case, one can use a quasiclassical approximation.\cite{Eilenberger68,Larkin68,KopninText} 

\begin{figure}
\includegraphics[width = 9cm]{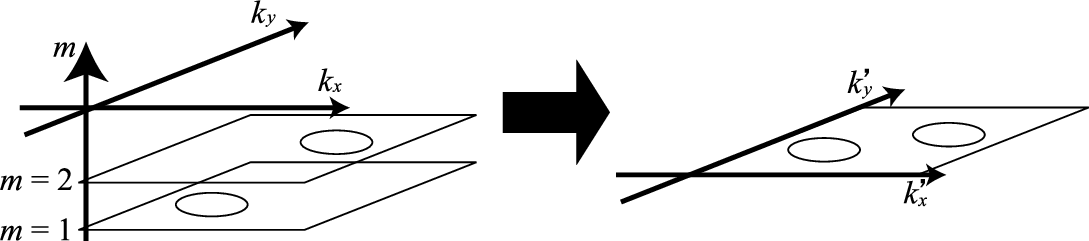}
\caption{\label{fig:andre}
Schematic figure of the effective Fermi surface.
}
\end{figure}

Since the Green function $\hat{G}_{0,mm}^{\rm b}(\Vec{k}_{1})$ is localized around the Fermi wave vector as a function of wave vector $\Vec{k}_{1 F}^{m}$
on the $m$-th band, 
the self-energy on the $i$-th band $\hat{\Sigma}_{i}^{\rm b, N}(\Vec{k}) \equiv \hat{\Sigma}_{ii}^{\rm b, N}(\Vec{k})$ 
is written as 
\begin{widetext}
\begin{equation}
\hat{\Sigma}_{i}^{\rm b, N}(\Vec{k}) \sim n_{\rm imp} \sum_{m} 
\int \frac{d S_{F,\hat{k}^{m}_{1}}
}{
(2\pi)^2 v_{F,\hat{k}^{m}_{1}}
}
\hat{v}_{i m}(\Vec{k},\hat{k}_{1}^{m}) \left( \int d \xi_{k_{1}^{m}}
\hat{G}_{0,mm}^{\rm b}(k_{1}^{m})
 \right) 
 \hat{v}_{mi}(\hat{k}_{1}^{m},\Vec{k}).
\end{equation}
\end{widetext}
$\hat{\Sigma}_{ii}^{\rm b, A}(\Vec{k})$ can be rewritten in a similar way.
Here $\hat{k}_{1}^{m}$ denotes the unit vector on the $m$-th band in the direction of $\Vec{k}_{1}^{m}$, 
$v_{F,\hat{k}^{m}_{1}}$ is the modulus of the Fermi velocity on the $m$-th band and 
$d S_{F,\hat{k}^{m}_{1}}$ is the Fermi-surface area element on the $m$-th band. 
We introduce the functions written as 
\begin{eqnarray}
\Delta(\Vec{k}_{F}) &\equiv& 
\sum_{m} \delta_{\Vec{k}_{F},\Vec{k}_{F}^{m}} \Delta_{m}(\Vec{k}^{m}_{F}), \\
v(\Vec{k}_{F},\Vec{k}_{F}') &\equiv& 
\sum_{m,m'} \delta_{\Vec{k}_{F},\Vec{k}_{F}^{m}} \delta_{\Vec{k}_{F}',\Vec{k}_{F}^{m'}}
\hat{v}_{mm'}(\Vec{k}_{F},\Vec{k}_{F}'),\: \:  \\
g(\Vec{k}_{F}) &\equiv& 
\sum_{m} \delta_{\Vec{k}_{F},\Vec{k}_{F}^{m}} \oint d \xi_{k^{m}} \hat{G}_{mm}^{\rm b}(k^{m}), \\
f(\Vec{k}_{F}) &\equiv& 
\sum_{m} \delta_{\Vec{k}_{F},\Vec{k}_{F}^{m}} \oint d \xi_{k^{m}} \hat{F}_{mm}^{\rm b}(k^{m}),
\end{eqnarray}
where the integral $\oint d\xi_{k^{m}}$ should be read as the contributions from poles close to the Fermi surface on 
the $m$-th band. 
With use of the above functions, 
the normal and anomalous parts of the self-energy are written, respectively, as 
\begin{eqnarray}
\Sigma^{\rm N}(\Vec{k}_{F}) &=& n_{\rm imp} \langle 
|v(\Vec{k}_{F},\Vec{k}_{1,F})|^2
g(\Vec{k}_{1,F})  
\rangle_{\rm FS_1}, \nonumber \\ \label{eq:bornself} \\
\Sigma^{\rm A}(\Vec{k}_{F}) &=& n_{\rm imp} \langle 
|v(\Vec{k}_{F},\Vec{k}_{1,F})|^2
f(\Vec{k}_{1,F})  
\rangle_{\rm FS_1}, \nonumber \\
\end{eqnarray}
where, 
\begin{eqnarray}
\langle A(\Vec{k}_{1,F}) \rangle_{\rm FS_1} &\equiv& 
\int \frac{d S_{F}(\Vec{k}_{1,F})}{(2\pi)^2 v_{F}(\Vec{k}_{1,F})} A(\Vec{k}_{1,F}) \\
&\equiv& 
 \sum_{m} 
\int \frac{d S_{F,\hat{k}^{m}_{1}}
}{(2\pi)^2 
v_{F,\hat{k}^{m}_{1}}
} A(\hat{k}_{1}^{m})
.
\end{eqnarray}
Here  
we introduce the effective single-band Fermi surface as shown in Fig.~\ref{fig:andre}, 
since the relation $\Vec{k}_{F}^{m} \neq \Vec{k}_{F}^{m'}$ is always satisfied for $m \neq m'$. 
Therefore,  
one can regard the $n$-band system as a disconnected single-band system 
omitting the band index $m$ and be allowed to translate an $n \times n$ matrix 
${\hat{ a}}$ into a scalar $a$ from now on. 

In the quasiclassical approximation, the self-energy can be determined as a local value $\Sigma(\Vec{r},\Vec{k}_{\rm F})$.\cite{KopninText}  
Therefore, one can calculate the self-energy by substituting $g(\Vec{r},\Vec{k}_{1})$ into Eq.~(\ref{eq:bornself}) 
in inhomogeneous systems. 

We introduce the quasiclassical Green function $\check{g}$ defined by 
\begin{eqnarray}
\check{g}(z,\Vec{r},\Vec{k}_{F})  \equiv \left(\begin{array}{cc}
g & f \\
- \tilde{f} & -g\end{array}\right), 
\end{eqnarray}
which is a $2 \times 2$ matrix in the Nambu space and is a function of 
complex frequency $z$, the Fermi wave vector $\Vec{k}_{F}$, a point $\Vec{r} = r (\cos \phi, \sin \phi)$ in 
real space. 
From now on, {\it check} $\check{a}$ denotes a $2 \times 2$ matrix in the Nambu space. 
The equation of motion for $\check{g}$ is written as 
\begin{equation}
- i \Vec{v}_{F}(\Vec{k}_{F}) \cdot \Vec{\nabla} \check{g} = \left[ 
z \check{\tau}_{3} - \check{\Delta}(\Vec{r},\Vec{k}_{F}) - \check{\Sigma},\check{g}
\right], \label{eq:eilen}
\end{equation}
supplemented by the normalization condition 
\begin{equation}
\check{g}^{2} = - \pi^{2} \check{1}. \label{eq:normal}
\end{equation}
Here $\check{\Delta}$ is given by 
\begin{eqnarray}
\check{\Delta}(\Vec{r},\Vec{k}_{F}) = 
\left(\begin{array}{cc}
0 & \Delta(\Vec{r},\Vec{k}_{F}) \\
- \Delta^{\ast}(\Vec{r},\Vec{k}_{F}) & 0\end{array}\right), 
\end{eqnarray}
and $\check{\Sigma}(z,\Vec{r},\Vec{k}_F)$ denotes the self-energy
\begin{eqnarray}
&&\check{\Sigma}(z,\Vec{r},\Vec{k}_{F})\nonumber\\
&&= n_{\rm imp} \langle 
|v(\Vec{k}_{F},\Vec{k}_{1,F})|^2
\check{g}(z,\Vec{r},\Vec{k}_{1,F})  
\rangle_{\rm FS_1}, \label{eq: checksigma}
\end{eqnarray}
and $\check{\tau}_{3}$ is a Pauli matrix in the Nambu space. 
In this paper, we consider the case where $\check{g}$ is an analytic function of $z$ in the upper half complex plane. 
Setting $z = \epsilon + i \delta$ with infinitesimal positive $\delta$, 
we obtain the retarded Green function. 
We use a special parameterization of the quasiclassical Green function 
to solve eq.~(\ref{eq:eilen}).\cite{Kato,Nagato,Higashitani, NagatoLow,SchopohlMaki,Schopohl} 
The solution $\check{g}$ of eq.~(\ref{eq:eilen}) can be written as 
\begin{eqnarray}
\check{g} = \frac{- i \pi}{1 + a b} 
\left(\begin{array}{cc}1- ab & 2 i a \\
- 2 i b & -(1 - a b)\end{array}\right). 
\label{eq: riccati}
\end{eqnarray}
Here $a$ and $b$ are the solutions of the following Riccati differential equations: 
\begin{eqnarray}
\Vec{v}_{F}(\Vec{k}_{F}) \cdot \Vec{\nabla} a &=& - 2(- i z  + i \Sigma_{11})a \nonumber \\
& &
- a^{2} (\Delta^{\ast} - \Sigma_{21}) + (\Delta + \Sigma_{12}), \label{eq:a} \\ 
\Vec{v}_{F}(\Vec{k}_{F}) \cdot \Vec{\nabla} b &=& + 2(- i z + i \Sigma_{11})b \nonumber \\
& & + b^{2} (\Delta + \Sigma_{12}) - (\Delta^{\ast} - \Sigma_{21}),\label{eq:b}
\end{eqnarray}
with $\check{\Sigma}=\{\Sigma_{ij}\}_{i,j=1,2}$. 
In the parameterization(\ref{eq: riccati}), the normalization condition (\ref{eq:normal}) 
is automatically satisfied. 
For simplicity, 
we solve the Riccati differential equations under a given form of pair potential. 
%

%
%
From now on, we drop the subscript $F$; e.g. $\Vec{v}_F \rightarrow \Vec{v}$ and $\Vec{k}_F \rightarrow \Vec{k}$. 
\section{Derivation of impurity scattering rate around a vortex }
\subsection{Kramer-Pesch approximation}
In this section, we explore low energy quasiparticle(QP)s around a single vortex and impurity effects within the scheme of quasiclassical theory\cite{Eilenberger68,Larkin68}.
Low energy QPs around a vortex have been discussed analytically and numerically\cite{Kramer74,Kopnin99,Kato,Kato03,Kato02,Tanuma09,Ichioka,SchopohlMaki,Schopohl,Eschrig,NagaiJPSJ,NagaiPRB}. In the absence of impurities, the low energy spectrum derived by Kramer and Pesch\cite{Kramer74} within the quasiclassical approximation essentially coincides with that of the Caroli-deGennes-Matricon mode\cite{Caroli64}. Even in impure and moderately clean superconductors, the contributions to the quasiclassical Green function from low energy QPs has been successfully worked out through a variant of the method used by Kramer and Pesch\cite{Kopnin99,Kato,Kato02}. From the pole of the quasiclassical Green function, we can obtain the spectrum and energy width of low energy QPs. In the quasiclassical theory, each QP has a definite momentum even in the presence of vortices. The energy width $\gamma_{\Vec{k}}$ of QP with momentum $\Vec{k}$ consists of the impurity scattering rate $\Gamma_{\Vec{k}\Vec{k'}}$ from the initial state (with momentum $\Vec{k}$) to the final state with momentum $\Vec{k'}$. From the expression for $\Gamma_{\Vec{k}\Vec{k'}}$, we can deduce the coherence factor in the impurity scattering between the Andreev bound states around a vortex. 

Equations~(\ref{eq:a}) and (\ref{eq:b}) contain $\Vec{\nabla}$ only through $\Vec{v} \cdot \Vec{\nabla}$ and hence these equations become one-dimensional problems on a straight line(quasiclassical trajectory) parallel to the Fermi velocity $\Vec{v}(\Vec{k})$ for a given momentum $\Vec{k}$. 
We denote by $\hat{\Vec{a}}$ and $\hat{\Vec{b}}$ the unit vectors along the crystal axes $X$ and $Y$, respectively. 
As proper coordinates to describe the quasiclassical trajectory, we introduce $s$ and $y$ by 
\begin{eqnarray}
\Vec{r} &=& X \hat{\Vec{a}} + Y \hat{\Vec{b}}, \\
&\equiv& s \hat{\Vec{v}} + y \hat{\Vec{u}}, 
\end{eqnarray}
with 
\begin{eqnarray}
\left(\begin{array}{c}\hat{\Vec{v}} \\
\hat{\Vec{u}}
\end{array}\right) 
\equiv 
\left(\begin{array}{cc}\cos \theta_{v} & \sin \theta_{v} \\
- \sin \theta_{v} & \cos \theta_{v}\end{array}\right)
\left(\begin{array}{c}\hat{\Vec{a}} \\
\hat{\Vec{b}}
\end{array}\right). 
\end{eqnarray}
Here $\theta_{v}$ is the angle between $\hat{\Vec{a}}$ and $\hat{\Vec{v}}$, the latter of which is the unit vector parallel to $\Vec{v}(\Vec{k})$. The symbol $s$ denotes the path along the trajectory and $y$ is the impact parameter of QP on the trajectory. We consider an axisymmetric single vortex located at $r=\sqrt{X^2+Y^2}=\sqrt{s^2+y^2}=0$ and take the pair potential of the form 
\begin{eqnarray}
\Delta(\Vec{k},\Vec{r})&=&  \Delta_{\infty}  d(\Vec{k}) f(r) \frac{X+i Y}{r}\nonumber\\
&=& \Delta_{\infty} d(\Vec{k}) f(r) e^{i \theta_{v}} \frac{s+i y}{r}.\label{eq: Deltakr}
\end{eqnarray}
Here $d(\Vec{k})$ describes the variation of the pair potential in momentum space (e.g., $d(\Vec{k})=1$ for $s$-wave superconductors and $d(\Vec{k})=k_x k_y/(k_x^2+k_y^2)$ for a $d$-wave superconductor). 
$\Delta_{\infty}|d(\Vec{k})|$ denotes the modulus of the pair-potential far away from the vortex ($\Delta_{\infty} > 0$). The function $f(r)$ describes the spatial variation of the modulus of the pair potential and satisfies $f(0) = 0$, $\lim_{r \rightarrow \infty} f(r) = 1$. . 

In the Kramer-Pesch approximation within the Riccati formalism in impure superconductors, 
we expand $a$ and $b$ in Eqs.(\ref{eq:a}) and (\ref{eq:b}) with respect to the impact parameter $y$, the complex frequency $z$ and the self-energy.  
Near a vortex, the superconducting pair potential 
up to with 
respect to $y$ can be written as 
\begin{equation}
\Delta(\Vec{k} ,\Vec{r})=\Delta_{0} + \Delta_{1}+{\cal O}(y^2)
\end{equation}
with 
\begin{equation}
\Delta_{0}=f(|s|) {\rm sign}(s)\Delta_{\infty} d(\Vec{k} ) e^{i \theta_{v}}, 
\quad\Delta_{1}=i \frac{y}{s} \Delta_{0}. 
\end{equation}
Following Refs.~\onlinecite{Kato} and \onlinecite{NagaiJPSJ}, 
we obtain $a$ and $b$ as 
\begin{eqnarray}
&&a=a_0+a_1+{\cal O}(z^2,y^2,\Sigma^2,z y,z\Sigma, y\Sigma)\nonumber\\
&&b=b_0+b_1+{\cal O}(z^2,y^2,\Sigma^2,z y,z\Sigma, y\Sigma)
\end{eqnarray}
with 
\begin{eqnarray}
a_{0} &=& - \: {\rm sign}(d(\Vec{k} )) e^{i \theta_{v}}, \\
b_{0} &=&  \: {\rm sign}(d(\Vec{k} )) e^{-i \theta_{v}}, \\
a_{1}&=& \frac{2i e^{u(s,\Vec{k})}}{|\Vec{v} (\Vec{k} )|}
\int_{- \infty}^{s} ds' \left( 
a_{0} (z - \tilde{\Sigma} ) + \frac{y}{s'}\Delta_{0} 
\right) e^{-u(s',\Vec{k})}, \nonumber \\ \\
 b_{1}&=& \frac{2i e^{u(s,\Vec{k})}}{|\Vec{v} (\Vec{k} )|}
\int_{\infty}^{s} ds' \left( 
- b_{0} (z - \tilde{\Sigma} ) + \frac{y}{s'}\Delta_{0}^{\dagger} 
\right) e^{-u(s',\Vec{k})}.\nonumber \\
\end{eqnarray}
Here we have introduced the following functions: 
\begin{eqnarray}
\tilde{\Sigma}(z,\Vec{r},\Vec{k} ) &\equiv& \Sigma_{11} - \frac{i}{2} \: {\rm sign}(d(\Vec{k} ))
\left( 
e^{i \theta_{v}} \Sigma_{21} + e^{- i \theta_{v}} \Sigma_{12}
\right), \: \: \: \: \: \: \:  \\
u(s,\Vec{k}) &=& \frac{2 |d(\Vec{k} )| \Delta_{\infty}}{|v (\Vec{k} )|}
\int_{0}^{|s|} ds' f(s').
\end{eqnarray}
At a small $|z|$ and a small impact parameter $y$, 
the Green function is thus written as\cite{Kato,NagaiJPSJ} 
\begin{eqnarray}
\check{g} &\sim& \frac{- 2 \pi i}{a_{1}b_{0} + a_{0} b_{1}} 
\check{M}(\Vec{k}),\nonumber\\
&\sim&\frac{\pi |v(\Vec{k})|\check{M}(\Vec{k})e^{-u(s,\Vec{k})}}{2C(\Vec{k})(z-E(y,\Vec{k})-\bar{\Sigma}(z,y,\Vec{k}))} 
\end{eqnarray}
with 
\begin{eqnarray}
&&\check{M}(\Vec{k}) \equiv \left(\begin{array}{cc}1 & i a_{0} \\
- i b_{0} & -1
\end{array}\right),\\
&&C(\Vec{k})=\int_0^\infty e^{-u(s,\Vec{k})}ds,\\
&&E(y,\Vec{k})=y F(\Vec{k}),\label{eq: E-y-k}\\
&&F(\Vec{k})=\frac{\Delta_\infty|d(\Vec{k})|}{C(\Vec{k})}\int_0^\infty \frac{f(s)}{s}e^{-u(s,\Vec{k})}ds,\label{eq: F-k}\\
&&\bar{\Sigma}(z,y,\Vec{k})\nonumber\\
&&=\frac{1}{2C(\Vec{k})}\int_{-\infty}^\infty \tilde{\Sigma}(z,s,y,\Vec{k})e^{-u(s,\Vec{k})}ds.\label{eq: Sigma-bar}
\end{eqnarray}
In the absence of impurities, $\bar{\Sigma}(z,y,\Vec{k})$ vanishes and the quasiclassical Green function has the pole at $z=E(y,\Vec{k})$, which is regarded as energy of QPs; correspondingly, the Andreev approximation of Bogoliubov-deGennes equation yields QP bound state with energy $E(y,\Vec{k})$, the wave function of which is localized near a vortex\cite{Volovik}. A schematic picture of an Andreev bound state in a vortex core is given, e.g. in Fig.~1 of Ref.~\onlinecite{Stone96} ($b$ and $x$ in Ref.~\onlinecite{Stone96} correspond, respectively, to $-y$ and $s$ in the present paper). Further we remark that $E(y,\Vec{k})$ for two-dimensional $s$-wave superconductors ($d(\Vec{k})=1$) reduces to the same form as that of Caroli-deGennes-Matricon mode\cite{Caroli64}(Recall that $y$ is the impact parameter with respect to the vortex center and hence the angular momentum $L$ parallel to $\hat{\Vec{a}}\times\hat{\Vec{b}}$ is given by $L=-k y$). 
The validity of the Kramer-Pesch approximation has been studied by Mel'nikov {\it et al.} by comparing with numerical results of the Bogoliubov-de Gennes equation\cite{Melnikov}. They showed that Kramer-Pesch approximation provides a reasonable description of the low-energy quasiparticle spectrum behavior when the radial profile of the pair-potential is given. 

Now we return to the system with impurities. Performing analytical continuation $z\rightarrow \epsilon+i\delta$, the quasiclassical Green function becomes
\begin{eqnarray}
&&\check{g}(\epsilon+i\delta,\Vec{k},\Vec{r})\nonumber\\
&&\sim\frac{\pi v(\Vec{k})\check{M}(\Vec{k})}{2C(\Vec{k})(\epsilon-E(y,\Vec{k})-E'(\epsilon,\Vec{k})+i \gamma(\epsilon,\Vec{k}))},
\end{eqnarray}
with
\begin{eqnarray}
&&E'(\epsilon,\Vec{k})={\rm Re}\bar{\Sigma}(\epsilon+i\delta,y=\epsilon/F(\Vec{k}),\Vec{k}),\nonumber\\
&&\gamma(\epsilon,\Vec{k})=-{\rm Im}\bar{\Sigma}(\epsilon+i\delta,y=\epsilon/F(\Vec{k}),\Vec{k})\ge 0.\label{eq: DeltaE-gamma}
\end{eqnarray}
$E'(\epsilon,\Vec{k})$ and $\gamma(\epsilon,\Vec{k})$ represent the QP energy shift and energy width due to impurities, respectively. We consider sufficiently clean superconductors so that the (non-selfconsistent) Born approximation is valid. For those systems, both $E'(\epsilon,\Vec{k})$ and $\gamma(\epsilon,\Vec{k})$ are small and $\check{g}$ is large when $\epsilon=y F(\Vec{k})$ is satisfied. We have thus fixed the value of $y$ to be $\epsilon/F(\Vec{k})$ in Eq.~(\ref{eq: DeltaE-gamma}). In the following, we ignore the effect of $E'(\epsilon,\Vec{k})$ and we discuss the energy width $\gamma(\epsilon,\Vec{k})$ in more detail. 

\subsection{Energy width and impurity scattering rate}
\begin{figure}
\includegraphics[width = 6cm]{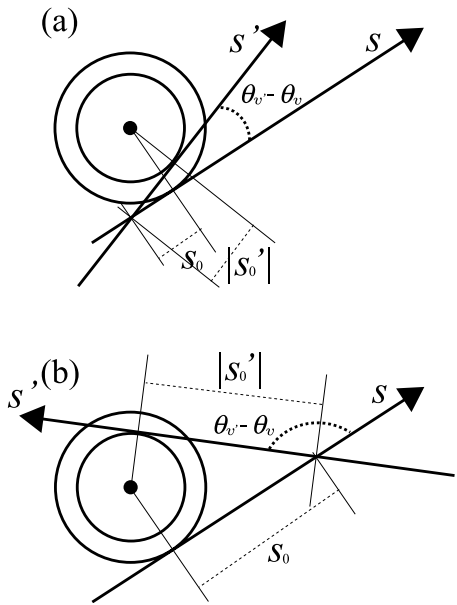}
\caption{\label{fig:scattering}
Schematic figures of (a) the forward and (b) the backward scatterings 
around a single vortex core in the low energy.
}
\end{figure}

In pure superconductors, 
the Green function near a vortex core at low energy can be written as 
\begin{equation}
\check{g}_{0}(\epsilon+i \delta, \Vec{r},\Vec{k}) \sim  \frac{\pi |\Vec{v} (\Vec{k} )|e^{-u(s,\Vec{k})}
}{
2C(\Vec{k} )
}
\frac{\epsilon - E(y,\Vec{k} ) - i \delta}{(\epsilon - E(y,\Vec{k} ))^{2} + \delta^{2}} \check{M}, \label{eq:kpa0}
\end{equation}
where $E(y,\Vec{k})$ has been given in Eqs.~(\ref{eq: E-y-k}) and (\ref{eq: F-k}). 
By substituting Eq.~(\ref{eq:kpa0}) into Eq.~(\ref{eq: checksigma}), 
the self-energy is written as 
\begin{widetext}
\begin{equation}
\Sigma_{ij}(\epsilon+i\delta,s,y,\Vec{k} ) = %
n_{\rm imp}\Bigl{\langle} |v_{\Vec{k} ,\Vec{k}_1 }|^2
 \frac{ \pi |\Vec{v} (\Vec{k}_1 )|  e^{- u(s'',\Vec{k}_1 )}}{2C(\Vec{k}_1 )} \frac{\epsilon - E(y',\Vec{k}_1 )- i\delta}{(\epsilon -E(y',\Vec{k}_1 ))^{2} +\delta^{2}} \check{M}_{ij}
\Bigl{\rangle}_{\rm FS_1}.
\end{equation}
Here we introduce the coordinates $(s'',y')$ in the direction of the momentum $\Vec{k}_1 $:
\begin{eqnarray}
y' &=& s \sin (\theta_{v} - \theta_{v'}) + y \cos (\theta_{v} - \theta_{v'}), \\
s'' &=& s \cos (\theta_{v} - \theta_{v'}) - y \sin (\theta_{v} - \theta_{v'}),
\end{eqnarray}
where $\theta_{v'}$ is the angle between $\hat{\Vec{a}}$ and the Fermi velocity $\Vec{v} (\Vec{k}_1 )$. 
With use of the above equation and $\delta \rightarrow 0$, ${\rm Im} \tilde{\Sigma}(\epsilon+ i\delta,s,y ,\Vec{k} )$ is written as 
\begin{align}
{\rm Im} \tilde{\Sigma}(\epsilon+i\delta,s,y,\Vec{k} ) &=
 {\rm Im}\: \Sigma_{11} - \frac{{\rm sign}\:(d(\Vec{k} ))}{2} \: {\rm Re} (e^{i \theta_{v}} \Sigma_{21} + e^{- i \theta_{v}} \Sigma_{12}) \\
&=-n_{\rm imp}\Bigl{\langle} |v_{\Vec{k} ,\Vec{k}_1 }|^2
 \frac{ \pi^{2} |\Vec{v} (\Vec{k}_1 )|  e^{- u(s'',\Vec{k}_1 )}}{2C(\Vec{k}_1 )}\tilde{M}(\Vec{k},\Vec{k}_1) 
 \delta(\epsilon - E(y',\Vec{k}_1 ))
\Bigl{\rangle}_{\rm FS_1},\label{eq: Im-tilde-Sigma}
\end{align}
with 
\begin{eqnarray}
\tilde{M}(\Vec{k},\Vec{k}_1)&=&M_{11} - \frac{{\rm sign}\:(d(\Vec{k} ))}{2} \: {\rm Re} (e^{i \theta_{v}} M_{21}(\Vec{k}_1) + e^{- i \theta_{v}} M_{12}(\Vec{k}_1))\nonumber\\
&=&1 - {\rm sign}[d(\Vec{k} )d(\Vec{k}_1 ))]\cos (\theta_{v} - \theta_{v'})\label{eq: Mtilde}.
\end{eqnarray}
Expression (\ref{eq: Mtilde}) is regarded as the coherence factor of the Andreev bound state around a vortex. 
From Eqs.~(\ref{eq: Sigma-bar}), (\ref{eq: DeltaE-gamma}) and (\ref{eq: Im-tilde-Sigma}), 
we obtain the expression for the energy width as
\begin{eqnarray}
\gamma(\epsilon,\Vec{k})&=&-\int_{-\infty}^\infty {\rm Im}\tilde{\Sigma}(\epsilon+i\delta,s,y=\epsilon/F(\Vec{k}),\Vec{k})e^{-u(s,\Vec{k})}ds\label{eq: gamma-epsilon-k}\\ 
&=&\frac{\pi^2 n_{\rm imp}}{4C(\Vec{k})} \left\langle\frac{
|v_{\Vec{k},\Vec{k}'}|^2\tilde{M}(\Vec{k},\Vec{k}')}{|\sin (\theta_{v} - \theta_{v'})| F(\Vec{k}') C(\Vec{k}')} 
e^{-u(s_{0},\Vec{k})}e^{-u(s_{0}',\Vec{k}')} \right\rangle_{\rm FS'},
\label{eq: gammafinal}
\end{eqnarray}
with 
\begin{eqnarray}
s_{0}(\Vec{k},\Vec{k}') &=& \frac{\epsilon}{\sin(\theta_{v}- \theta_{v'})} \left( 
\frac{1}{F(\Vec{k}')}- \frac{\cos(\theta_v-\theta_{v'})}{F(\Vec{k})}\right), \\
s_{0}'(\Vec{k},\Vec{k}') &=& s_{0}(\Vec{k},\Vec{k}') \cos (\theta_{v} - \theta_{v'}) - \frac{\epsilon}{F(\Vec{k})}\sin(\theta_{v} - \theta_{v'}), \\
\langle \cdots \rangle_{\rm FS'} &\equiv& 
\int \cdots \frac{d S (\Vec{k}')}{(2\pi)^2 v (\Vec{k}')}. 
\end{eqnarray}
From (\ref{eq: gamma-epsilon-k}) to (\ref{eq: gammafinal}), we have exchanged the order of the $s$-integration and $\langle \cdots \rangle_{\rm FS'}$ and performed the $s$-integration. By expressing Eq.~(\ref{eq: gammafinal}) as
\begin{equation}
\gamma(\epsilon,\Vec{k})=\int \frac{d S (\Vec{k}')}{(2\pi)^2 v (\Vec{k}')}\Gamma_{\Vec{k},\Vec{k}'}(\epsilon)
\end{equation}
with 
\begin{equation}
\Gamma_{\Vec{k},\Vec{k}'}(\epsilon) = \frac{\pi^2 n_{\rm imp}|v_{\Vec{k},\Vec{k}'}|^2 \tilde{M}(\Vec{k},\Vec{k}')v(\Vec{k}')}{4|\sin (\theta_{v} - \theta_{v'})|C(\Vec{k})F(\Vec{k}')C(\Vec{k}')} 
e^{-u(s_{0},\Vec{k})}e^{-u(s_{0}',\Vec{k}')},\label{eq:gamma}
\end{equation}
we read $\Gamma_{\Vec{k},\Vec{k}'}(\epsilon)$ as the impurity scattering rate  for the quasiparticles from the initial state $\Vec{k}$ to the final state $\Vec{k}'$ in a vortex core. With use of the impurity scattering rate in the normal state, 
\begin{equation}
\Gamma_{\Vec{k},\Vec{k}'}(\epsilon)=\pi n_{\rm imp}
|v_{\Vec{k},\Vec{k}'}|^2\equiv\Gamma^{\rm N}_{\Vec{k},\Vec{k}'}
\end{equation}
we introduce the normalized impurity scattering rate
\begin{equation}
\tilde{\Gamma}_{\Vec{k},\Vec{k}'}(\epsilon)=
\frac{\Gamma_{\Vec{k},\Vec{k}'}(\epsilon)}{\Gamma^{\rm N}_{\Vec{k},\Vec{k}'}}=\frac{\pi \tilde{M}(\Vec{k},\Vec{k}')v(\Vec{k}')}{4|\sin (\theta_{v} - \theta_{v'})|C(\Vec{k})F(\Vec{k}')C(\Vec{k}')} 
e^{-u(s_{0},\Vec{k})}e^{-u(s_{0}',\Vec{k}')}.\label{eq:Gamma-normal}
\end{equation}
\end{widetext}
For superconductors with $|d(\Vec{k})|\sim 1$ and $|\Vec{v}(\Vec{k})|\sim v$ isotropic or nearly isotropic on the Fermi surfaces, $C(\Vec{k})\sim v/\Delta_\infty$ and $F(\Vec{k})\sim v$. Hence the dependence on momentum transfer of the normalized impurity scattering rate (\ref{eq:Gamma-normal}) mainly depends on the scattering angle $(\theta_{v} - \theta_{v'})$ 
and the sign-change of pair-potential ${\rm sign}[d(\Vec{k})d(\Vec{k}')]$ (which QPs feel during the scattering process). 
The normalized impurity scattering rate (\ref{eq:Gamma-normal}) also depends on 
the spatial distribution of the wavefunction $e^{-u(s_{0},\Vec{k})}$ for QP with momentum $\Vec{k}$ along the quasiclassical paths and that with $\Vec{k}'$ . 
One can clearly define the trajectory for the quasiparticles forming the Andreev bound states 
as shown in Fig.~\ref{fig:scattering}. 
Therefore, 
we can discuss 
the impurity scattering in vortex cores, in detail as presented in the next section.  

For superconductors with strongly anisotropic $|d(\Vec{k})|$ or $|\Vec{v}(\Vec{k})|$ on the Fermi surfaces (e.g. a $d$-wave superconductor), the momentum dependences of $C(\Vec{k})$ and $F(\Vec{k})$ also affect the dependence of momentum transfer of the normalized impurity scattering rate.

\section{Preliminary analysis}
\subsection{Sign-conserved scattering}
We consider the case where $d(\Vec{k})d(\Vec{k}') > 0$. 
In this case, the sign of the pair potential does not change during the scattering process. 
When the momentum dependence of the impurity scattering matrix $|v_{\Vec{k},\Vec{k}'}|^2 $ is 
not so large, the scattering angle $(\theta_{v} - \theta_{v'})$ is the most important factor for the impurity scattering rate. 
The dependence on the scattering angle $(\theta_{v} - \theta_{v'})$ in Eq.~(\ref{eq:gamma}) can be 
written as 
\begin{equation}
\tilde{\Gamma}_{\Vec{k},\Vec{k}'} \propto \frac{
 1- \cos (\theta_{v} - \theta_{v'})
 }
{
|\sin (\theta_{v} - \theta_{v'})|
}
{\rm e}^{-u(s_0,\Vec{k})}{\rm e}^{-u(s'_0,\Vec{k}')}.
\end{equation}
One can find that the forward scattering ($\theta_{v} - \theta_{v'} = 0$)
is suppressed since the coherence factor becomes zero 
\begin{equation}
\frac{ 1- \cos (\theta_{v} - \theta_{v'})}{
|\sin (\theta_{v} - \theta_{v'})|} \sim 
|\theta_{v} - \theta_{v'}|.
\end{equation}
In the case of the backward scattering ($\theta_{v} - \theta_{v'} = \pi$), 
in spite of the large coherence factor, the scattering rate is small 
since the point of the scattering is far from the vortex core 
so that the amplitude of the wavefunction of QPs $\exp(-u(s_0,\Vec{k}))\exp(-u(s_0,\Vec{k}_0))$ becomes small with large $s_0$ and $s_0'$ (see Fig.~\ref{fig:scattering}(b)).

\subsection{Sign-reversed scattering}
In the case where $d(\Vec{k})d(\Vec{k}') < 0$, 
the dependence on the scattering angle $(\theta_{v} - \theta_{v'})$ in Eq.~(\ref{eq:gamma}) can be 
written as 
\begin{equation}
\tilde{\Gamma}_{\Vec{k},\Vec{k}'} \propto \frac{
 1+ \cos (\theta_{v} - \theta_{v'})
 }
{
|\sin (\theta_{v} - \theta_{v'})|
}{\rm e}^{-u(s_0,\Vec{k})}{\rm e}^{-u(s'_0,\Vec{k}')}. \label{eq:sgr}
\end{equation}
The backward scattering ($\theta_{v} - \theta_{v'} = \pi$) in the present case is suppressed by both the coherence factor and the amplitudes of QP wavefunctions. 
When the impact parameters $y$ and $y'$ are equal (i.e. when the radii of two circles are equal in Fig.~\ref{fig:scattering}(a)), the forward scattering becomes large 
since the scattering point is near a vortex core ($\theta_v-\theta_{v'}\rightarrow 0$, $s_0,s_0'\rightarrow 0$)(see Fig.~\ref{fig:scattering}(a)). 
When the impact parameters $y$ and $y'$ are nearly equal, the scattering rate becomes large at a small but finite angle where the scattering point is 
near a vortex core. 

It should be noted that these sign-reversed forward (or nearly forward) scatterings {\it hardly} occur in single-band 
superconductors because quasiparticles close to each other in momentum space 
feel the same sign of the pair potential.  
In multi-band superconductors with electron and hole circular-like Fermi surfaces such as Fe-based materials, 
the inter-band scatterings can become the sign-reversed forward scatterings. 
Let $\Vec{q}$ denote the momentum transfer $\Vec{q} \equiv \Vec{k}' - \Vec{k}$. The $\Vec{q}$-dependence of the impurity scattering rate is then anomalous  
in sign-reversing $s$-wave superconductors 
since the intensity of the sign-reversed forward scatterings is much larger than that of all other scatterings. 
%
%
\section{Iron-based superconductors}
\subsection{Model}
\begin{figure}
\includegraphics[width = 7cm]{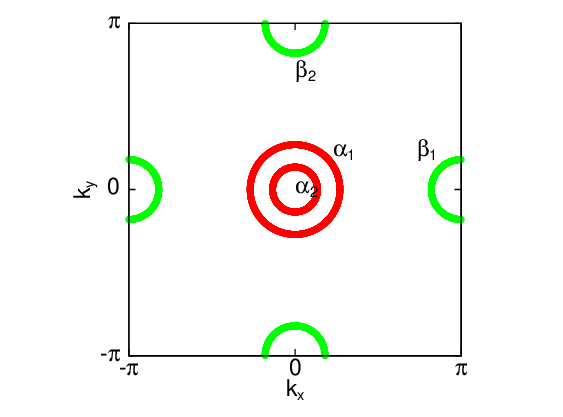}
\caption{\label{fig:fermi}
(Color online) 
Two hole Fermi surfaces $\alpha_{1},\alpha_{2}$ and two electron Fermi surfaces $\beta_{1},\beta_{2}$. 
}
\end{figure}

We show the $\Vec{q}$-dependence of the impurity scattering rate $\tilde{\Gamma}
_{\Vec{k},\Vec{k+q}}$ 
in two-band superconductors as simplified models for the Fe-based superconductors.  
We consider the system with two hole Fermi surfaces and two electron Fermi surfaces 
as shown in Fig.~\ref{fig:fermi} 
since the Fe-based superconductors have the multiple Fermi surfaces.\cite{DingEPL} 
The hole Fermi surface $\alpha_{1}$ ($\alpha_{2}$) 
is located around $(k_{x},k_{y}) = (0,0)$ with the diameter $1.2/\sqrt{2}$ ($0.6/\sqrt{2}$) 
and the electron Fermi surface $\beta_{1}$ ($\beta_{2}$) is located around 
$(k_{x},k_{y}) = (\pi,0)$ ($(k_{x},k_{y}) = (0,\pi)$) with the diameter $0.8/\sqrt{2}$. 

For simplicity, the amplitude of the Fermi velocity is 
taken to be isotropic on each Fermi surface in momentum space. 
We set $|\Vec{v}(\Vec{k})| =v$ on the hole Fermi surfaces $\alpha_{1}$, $\alpha_{2}$ and $|\Vec{v}(\Vec{k})| =0.99 v$ on the electron Fermi surfaces $\beta_{1}$, $\beta_{2}$.   
We set the radial profile $f(r)$ of the pair-potential as $\tanh(r/\xi)$ with the coherence length $\xi=v/(\pi\Delta_\infty)$.
We consider normalized scattering rate 
$
\tilde{\Gamma}_{\Vec{k},\Vec{k+q}}$ 
for various superconducting pairing symmetries; the isotropic $s$-wave $d(\Vec{k}) = 1$, 
the $d$-wave $d(\Vec{k}) = -(k_{x}^{2}-k_{y}^{2})/(k_{x}^{2}+k_{y}^{2})$ 
and the isotropic $\pm s$ wave $d(\Vec{k}) = -1$ for $\alpha$ bands and $d(\Vec{k}) = 1$ for $\beta$ bands. 
\begin{figure}
  \begin{center}
    \begin{tabular}{p{40mm}p{40mm}}
      \resizebox{40mm}{!}{\includegraphics{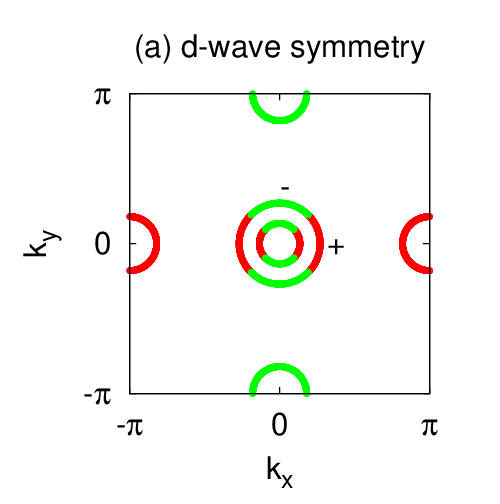}} &
      \resizebox{40mm}{!}{\includegraphics{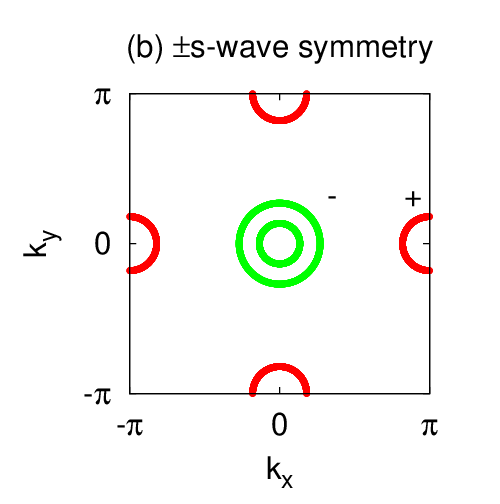}} 
    \end{tabular}
\caption{\label{fig:gap}
(Color online) 
Schematic figures of the sign of the pair functions for (a):$d$-wave and (b):$\pm s$-wave symmetries.
}
  \end{center}
\end{figure}
\subsection{Isotropic $s$-wave case}
\begin{figure}
  \begin{center}
    \begin{tabular}{p{50mm}p{30mm}}
      \resizebox{50mm}{!}{\includegraphics{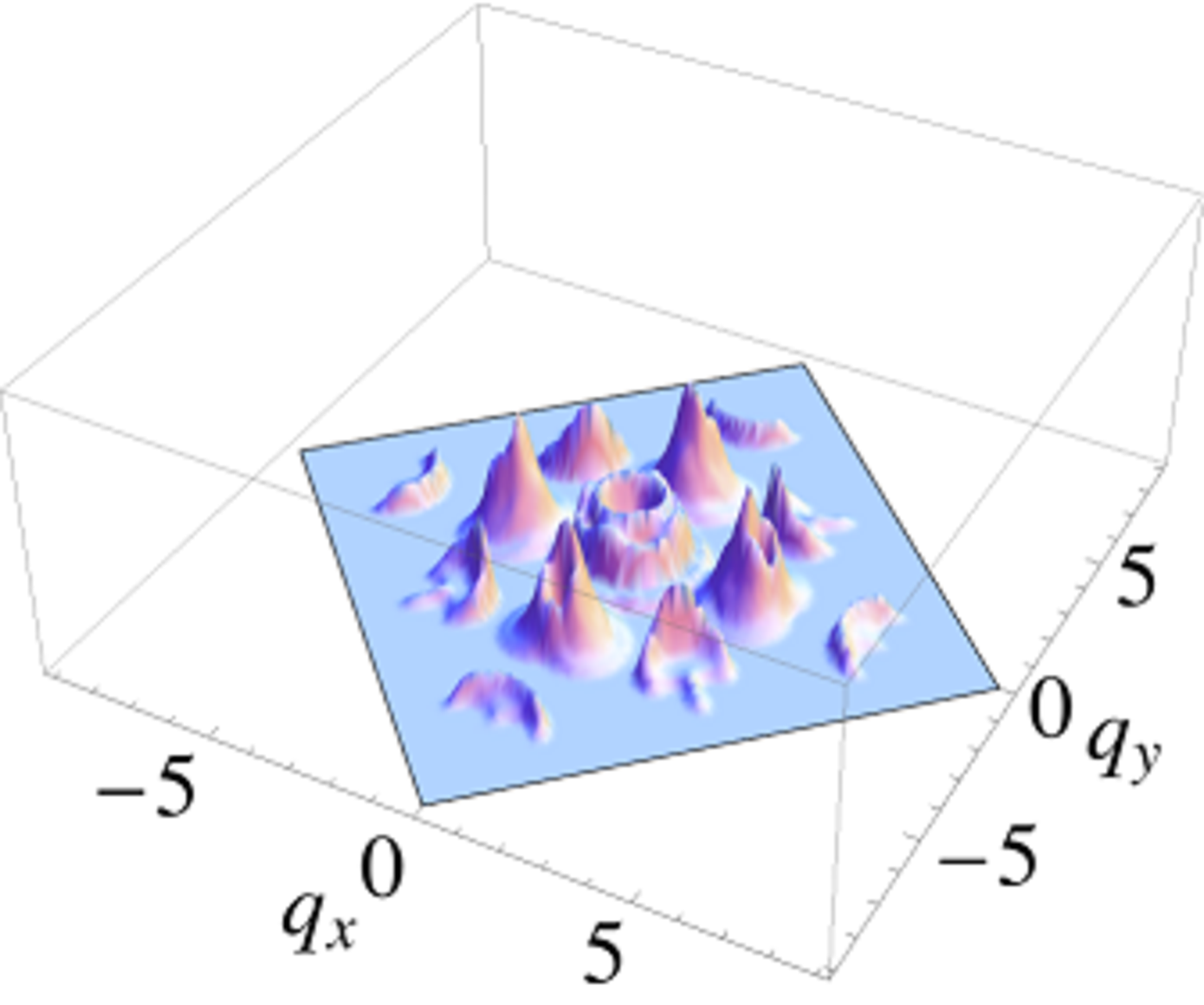}} &
      \resizebox{30mm}{!}{\includegraphics{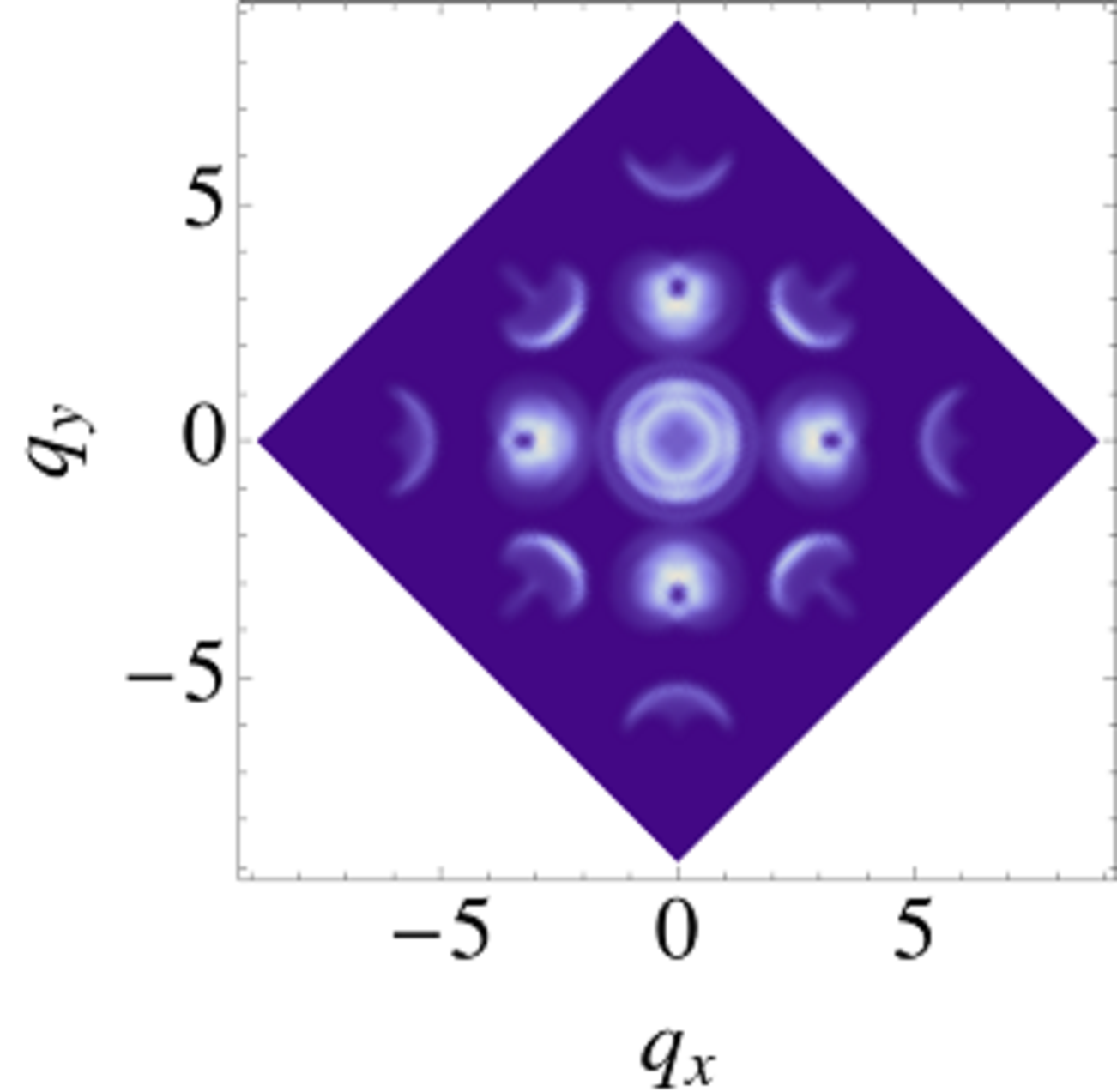}} 
    \end{tabular}
\caption{\label{fig:zuspp}
(Color online) 
$\Vec{q}$-dependence of the normalized impurity scattering rate $\tilde{\Gamma}_{\Vec{k},\Vec{k+q}}$ in the isotropic $s$-wave superconductors. 
The energy is $\epsilon= 0.3 \Delta_{\infty}$.
}
  \end{center}
\end{figure}

First, we consider the isotropic $s$-wave superconductivity. 
In this case, all scatterings are sign-conserved scatterings. 
As shown in Fig.~\ref{fig:zuspp}, 
the normalized impurity scattering rate 
is finite everywhere 
in $\Vec{q}$-space. 
This result is consistent with the result in Sec.~IV.~(A).

\begin{figure}
  \begin{center}
    \begin{tabular}{p{50mm}p{30mm}}
      \resizebox{50mm}{!}{\includegraphics{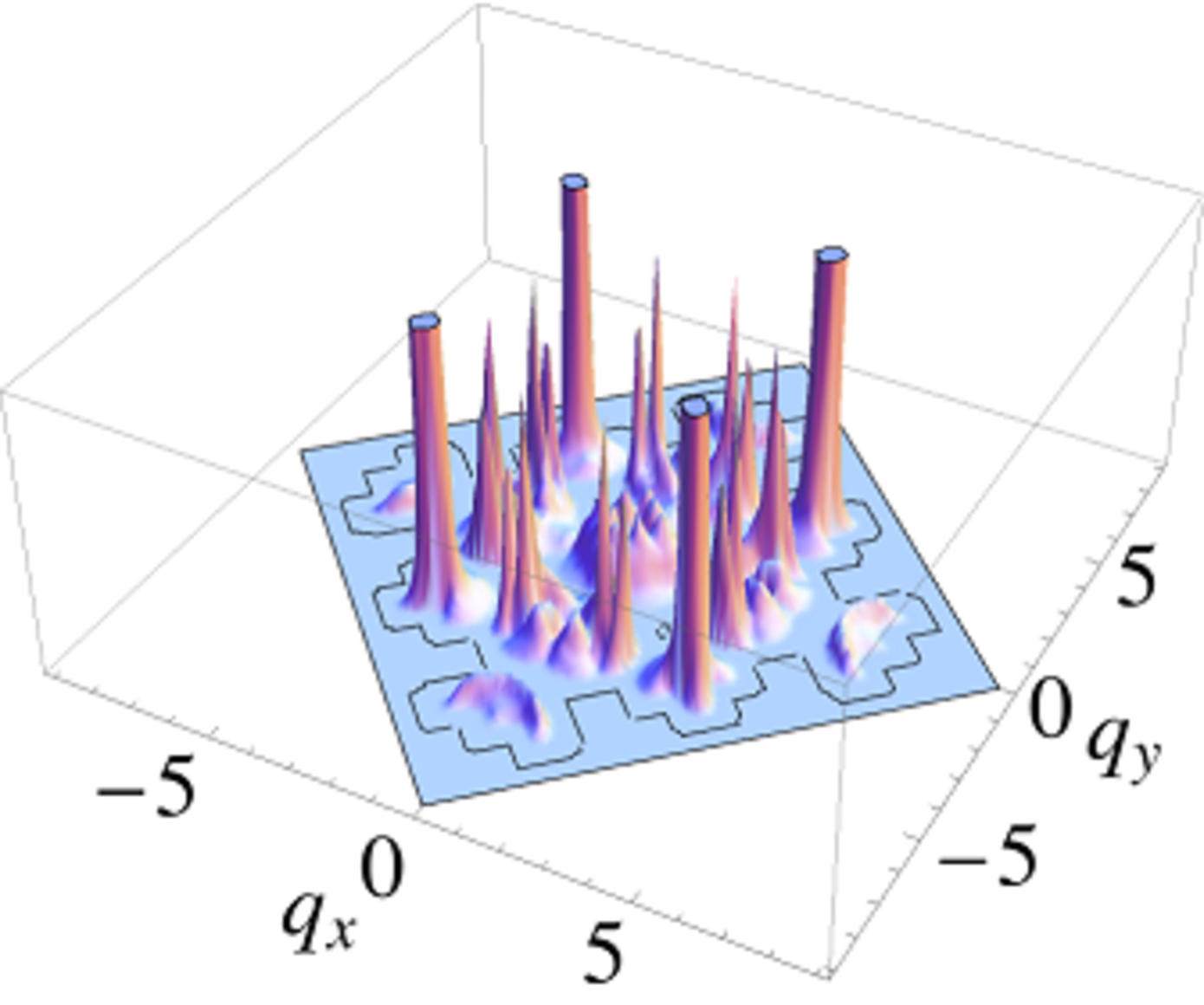}} &
      \resizebox{30mm}{!}{\includegraphics{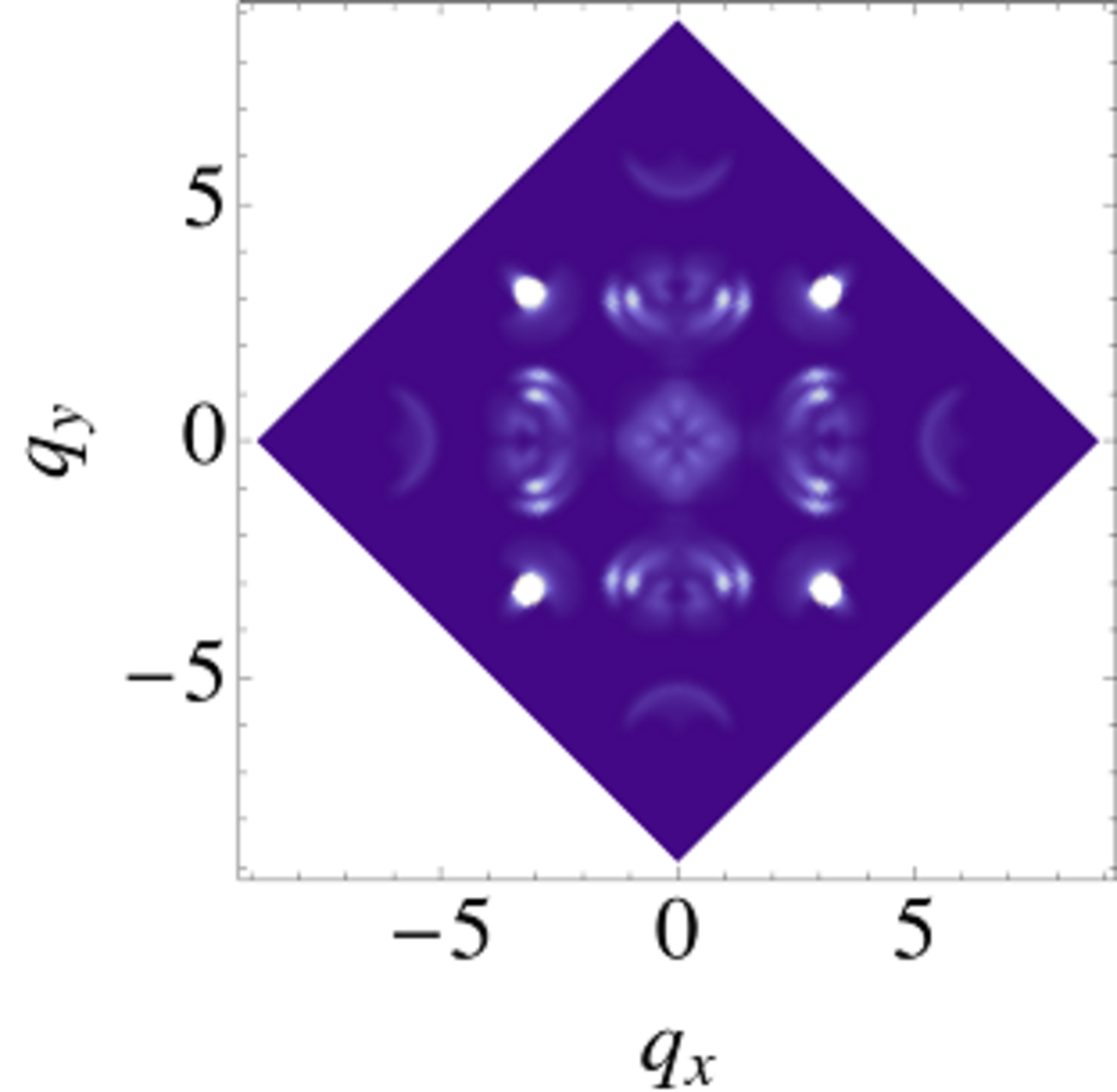}} 
    \end{tabular}
\caption{\label{fig:zudw}
(Color online)
$\Vec{q}$-dependence of the normalized impurity scattering rate $\tilde{\Gamma}_{\Vec{k},\Vec{k+q}}$ in the $d$-wave superconductors $d(\Vec{k})= 
 -(k_{x}^{2}- k_{y}^{2})/(k_{x}^{2}+k_{y}^{2})$. 
The energy is $\epsilon= 0.3 \Delta_{\infty}$.
}
  \end{center}
\end{figure}
\subsection{$d$-wave case}
\begin{figure}
\includegraphics[width = 6cm]{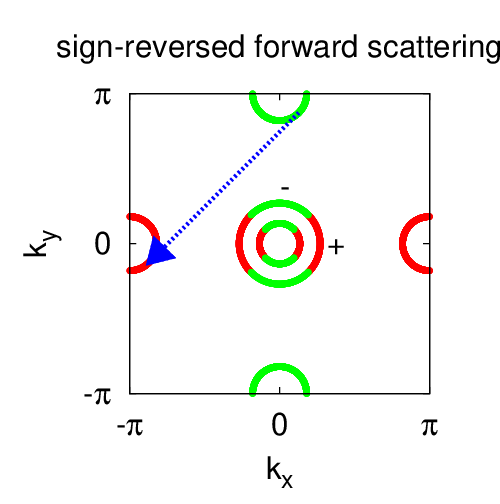}
\caption{\label{fig:fig4c}
(Color online) 
The arrow denotes the sign-reversed forward scattering in $d$-wave case. 
}
\end{figure}

Next, we consider the $d$-wave superconductivity as shown in Fig.~\ref{fig:gap}(a). 
In this case, the scatterings between $\beta_{1}$ and $\beta_{2}$ Fermi surfaces are dominant 
since some of these scatterings are the sign-reversed forward scatterings. 
More specifically, the scattering from $(k_{x},k_{y}) = (0.4,\pi-0.4)$ ($\theta_{v} = 0$) 
on the $\beta_{2}$ Fermi surface 
 to $(k_{x},k_{y}) = (\pi-0.4,0.4)$ ($\theta_{v'} = 0$) on the $\beta_{1}$ Fermi surface 
is the sign-reversed forward scattering as shown in Fig.~\ref{fig:fig4c}. 
Recall that 
we define the forward scatterings as the ones satisfying $\theta_{v'} \sim \theta_{v}$. 
From Eq.~(\ref{eq:sgr}), one see that the impurity scattering rate diverges for this scattering. 
It should be noted that the condition for these sign-reversed forward scatterings to exist 
is sensitive to the shape of the Fermi surface and/or the direction of the Fermi velocity 
in the $d$-wave superconductors. 

\begin{figure}
  \begin{center}
    \begin{tabular}{p{50mm}p{30mm}}
      \resizebox{50mm}{!}{\includegraphics{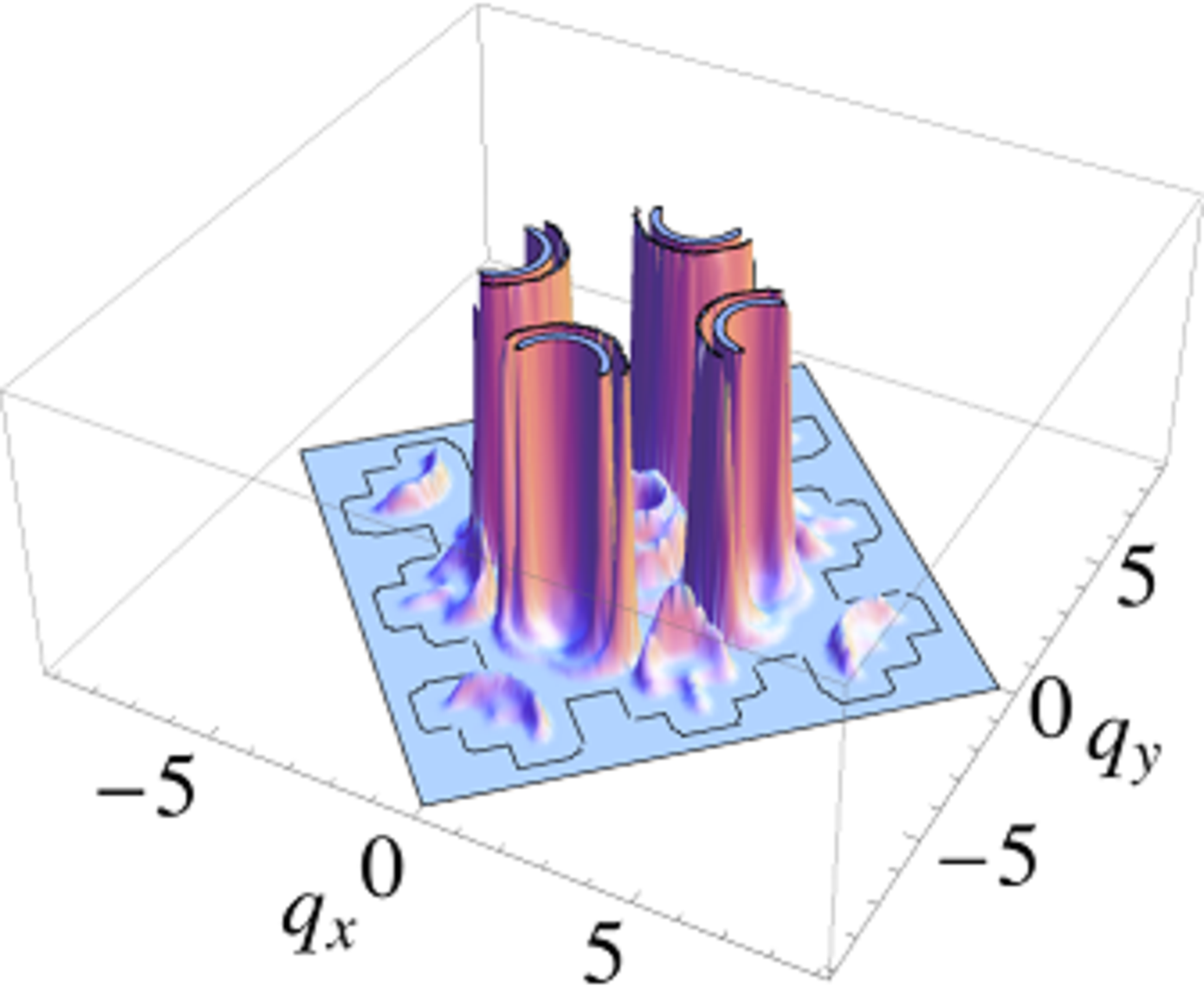}} &
      \resizebox{30mm}{!}{\includegraphics{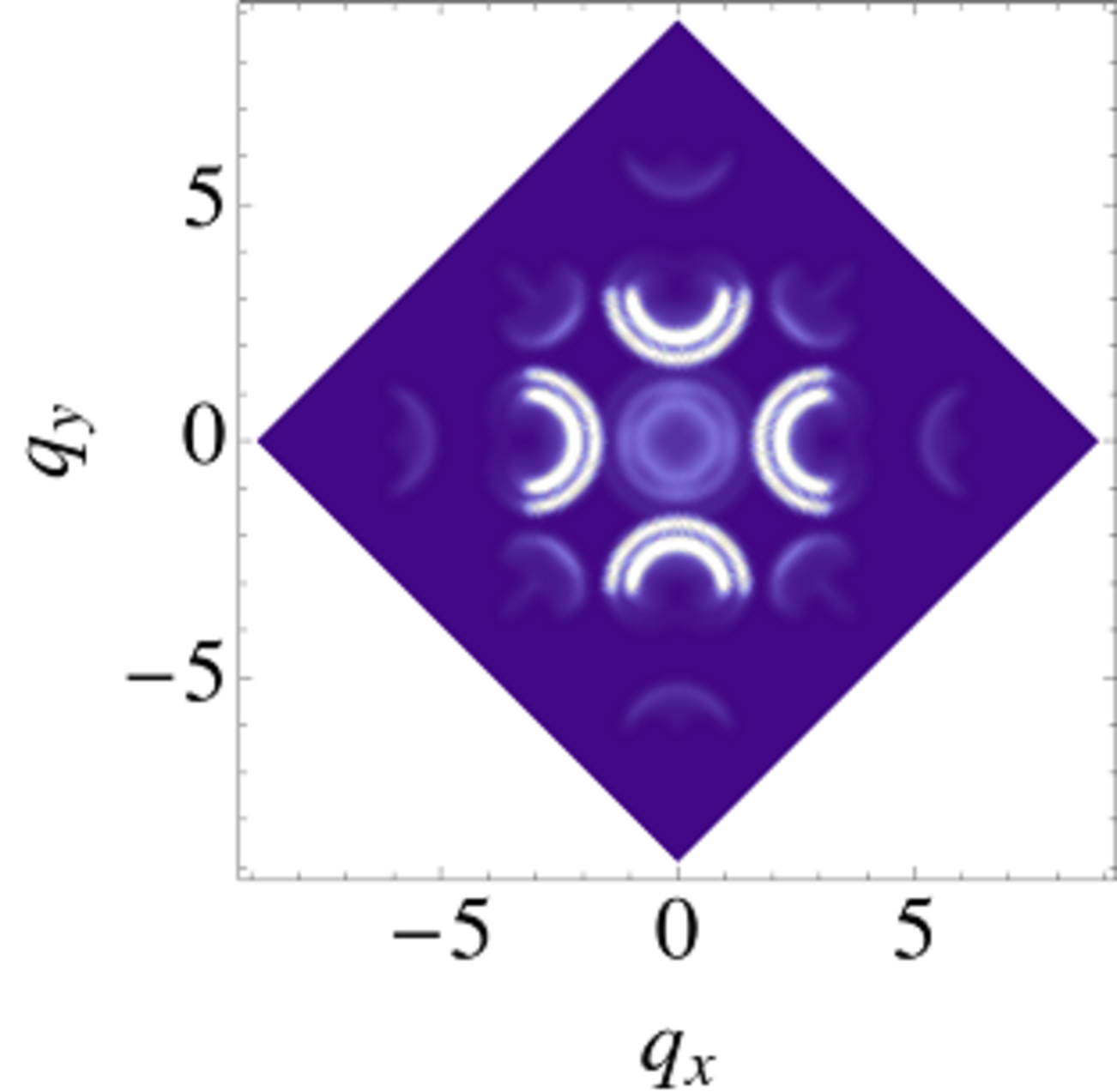}} 
    \end{tabular}
\caption{\label{fig:zuspm}
(Color online)
$\Vec{q}$-dependence of the normalized impurity scattering rate $\tilde{\Gamma}_{\Vec{k},\Vec{k+q}}$ in the isotropic $\pm s$-wave superconductors. 
The energy is $\epsilon= 0.3 \Delta_{\infty}$.
}
  \end{center}
\end{figure}
\subsection{Isotropic $\pm s$-wave case}
Finally, we consider the isotropic $\pm s$-wave superconductivity as shown in Fig.~\ref{fig:gap}(b). 
In this case, the inter-band scatterings are dominant. 
As shown in Fig.~\ref{fig:zuspm}, there are the arc-like strong intensity distributions $\tilde{\Gamma}_{\Vec{k},\Vec{k}+\Vec{q}}$ in $\Vec{q}$-space. 
These intensity distributions are caused by the sign-reversed forward scatterings from the arc-like regions on the $\alpha_{1}$ 
or $\alpha_{2}$ Fermi surfaces to those on the $\beta_{1}$ or $\beta_{2}$ Fermi surfaces. 
The regions where the sign-reversed forward scatterings occur for the $\pm s$-wave superconductors are broader 
than those for the $d$-wave superconductors. 
These properties are robust in the $\pm s$-wave superconductors which has two or more Fermi surfaces as we assume. 
As shown in Eq.~(\ref{eq:sgr}), one of the most important 
factors for the sign-reversed scatterings 
is the relation of the directions of the Fermi velocity of the quasiparticles before and after scatterings. 
In the systems with electron and hole Fermi surfaces such as the Fe-based superconductors, 
the more the shapes of the Fermi surfaces become similar to each other, 
the regions of the strong intensity become broader in $\Vec{q}$-space. 

\section{Discussion}
We discuss the energy dependence of the energy width $\gamma(\epsilon,\Vec{k})$ for the system with the Fermi surfaces shown in Fig.~\ref{fig:fermi}. 
Let $n$ ($= 1,2,3,4$) be the index of the disconnected Fermi surface ($\alpha_{1}$, $\alpha_{2}$, $\beta_{1}$, $\beta_{2}$).
We take the modulus of the Fermi velocity and $d(\Vec{k})$ to be, respectively, constants $v_n$ and $d_n$ within each Fermi surface.  The energy width then depends only on $\epsilon$ and $n$. Further we approximate $|v_{\Vec{k},\Vec{k}'}|^2$ by the average $\overline{|v_{\Vec{k},\Vec{k}'}|^2}$ over the Fermi surfaces, for simplicity.  
By performing the Fermi-surface-integration before the $s$-integration, 
the energy width $\gamma(\epsilon,\Vec{k})$ can then be written as
\begin{widetext}
\begin{eqnarray}
\gamma(\epsilon,\Vec{k})=\gamma(\epsilon,n)&=&
\frac{n_{\rm imp}\overline{|v_{\Vec{k}\Vec{k}'}|^2}}{4C_n}\sum_{n'}\frac{k_{n'} v_{n'}}{C_{n'}F_{n'}}\int_0^\infty\frac{{\rm Re}s'}{|s'|^2}e^{-u(s',n')}e^{-u(s,n)}{\cal M}_{n,n'}(\epsilon,s)ds,\label{eq: gammaepsilonk}
\end{eqnarray}
with 
\begin{eqnarray}
{\cal M}_{n,n'}(\epsilon,s)=
\delta_{n,n'}\frac{s^2 F_n^2}{s^2F_n^2+\epsilon^2}+(1-\delta_{n,n'})\left(1-\frac{{\rm sign}(d_n d_{n'})\epsilon^2 F_{n} F_{n'}^{-1}}{s^2 F_n^{2}+\epsilon^2}\right)
\quad
s' \equiv \sqrt{s^{2} + \epsilon^{2} (F_{n}^{-2} - F_{n'}^{-2})}. 
\label{eq: Mnnprime}
\end{eqnarray}
In Eq.~(\ref{eq: gammaepsilonk}), $k_{n}$ denotes the radius of the $n$-th Fermi surface. 

Introducing the energy width in the normal state
\begin{eqnarray}
\gamma^{\rm N}&=&\int\frac{d S_{\Vec{k}'}}{(2\pi)^2 v(\Vec{k}')}\Gamma^{\rm N}(\Vec{k},\Vec{k}')\sim\frac{n_{\rm imp}\overline{|v_{\Vec{k},\Vec{k}'}|^2}}{2}\sum_{n'}\frac{k_{n'}}{v_{n'}},
\end{eqnarray}
we obtain the normalized energy width $\tilde{\gamma}(\epsilon,n)$
\begin{eqnarray}
\tilde{\gamma}(\epsilon,n)\equiv\frac{\gamma(\epsilon,n)}{\gamma^{\rm N}}&=&
\left(2C_n\sum_{n''}\frac{k_{n''}}{v_n''}\right)^{-1}\sum_{n'}\frac{k_{n'} v_{n'}}{C_{n'}F_{n'}}\int_0^\infty\frac{{\rm Re}s'}{|s'|^2}e^{-u(s',n')}e^{-u(s,n)}{\cal M}_{n,n'}(\epsilon,s)ds.
\label{eq: tildegamma}
\end{eqnarray}
\end{widetext}

From Eq.~(\ref{eq: Mnnprime}), we see that the integral in Eq.~(\ref{eq: tildegamma}) diverges logarithmically when $\epsilon=0$ or 
\begin{equation}
d_n d_{n'}<0\quad \mbox{and}\quad F_n=F_{n'},\quad\mbox{for }n\ne n'.
\label{eq: dndnnegative}
\end{equation}
The divergence coming from Eq.~(\ref{eq: dndnnegative}) can be removed by introducing a slight difference of $d_n$ or $v_n$ among different Fermi surfaces. 
For numerical evaluation of $\tilde{\gamma}(\epsilon,n)$, we take 
$d_{1,2} = -1$, $d_{3,4} = 1$, 
$v_{1} = v_{2}=v$, $v_{3} = v_{4}=1.25v$, $\xi=v/(\pi \Delta_\infty)$ and $f(r)=\tanh(r/\xi)$.  

\begin{figure}
\includegraphics[width = 7cm]{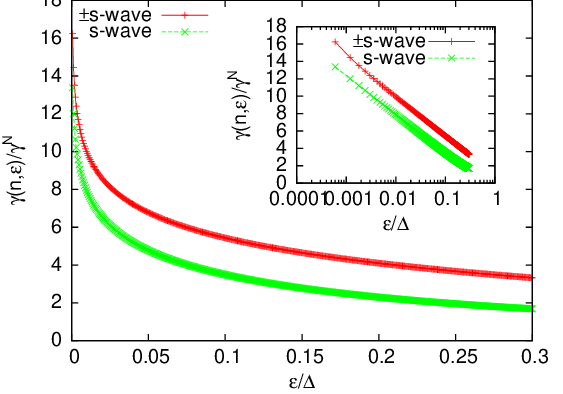}
\caption{\label{fig:fig8}
(Color online) Energy dependence of the normalized energy width $\tilde{\gamma}(\epsilon,n)$ with $n=4$ in the isotropic $\pm s$-wave and $s$-wave superconductors. 
}
\end{figure}

As shown in Fig.~\ref{fig:fig8}, the normalized energy width $\tilde{\gamma}(\epsilon,n)$ with $n=4$ has logarithmic energy dependence, which has been found also in the single-band $s$-wave case within the Born approximation\cite{KopninLopatin}. 
The energy dependence of the normalized energy width $\tilde{\gamma}(\epsilon,n)$ with $n=1,2,3$ is 
 similar to this. 
We note that these results do not change in the case of $s$-wave superconductors as shown in Fig.~\ref{fig:fig8}, qualitatively.
With use of Fig.~\ref{fig:fig8}, we discuss the validity of our results. The non-selfconsistent Born approximation is valid when $\gamma(\epsilon,n)/\epsilon\ll 1$, which is rewritten as 
\begin{equation}
\tilde{\gamma}(\epsilon,n) \left(\frac{\gamma^{\rm N}}{\Delta_\infty}\right)/\left(\frac{\epsilon}{\Delta_\infty}\right)\ll 1.\label{eq: Borncondition}
\end{equation}
On the other hand, the Kramer-Pesch approximation is valid when $\epsilon/\Delta_\infty\ll 1$.
The ratio $\gamma^{\rm N}/\Delta_\infty$ is, roughly speaking, the same magnitude of $\xi/l$ with the mean free path $l$ in the normal state. 
When $\gamma^{\rm N}/\Delta_\infty\sim 0.1$, we see from Fig.~\ref{fig:fig8} that the validity conditions of non-selfconsistent Born and Kramer-Pesch approximations hold when 
\begin{equation}
0.07\ll \frac{\epsilon}{\Delta_\infty}\ll 1.
\end{equation}
When $\gamma^{\rm N}/\Delta_\infty\sim 0.01$, on the other hand, our result is valid  when 
\begin{equation}
0.01\ll \frac{\epsilon}{\Delta_\infty}\ll 1.
\end{equation}
Within the energy regions evaluated above, we can safely discuss the physical implication of our results.  
\begin{figure}
\includegraphics[width = 7cm]{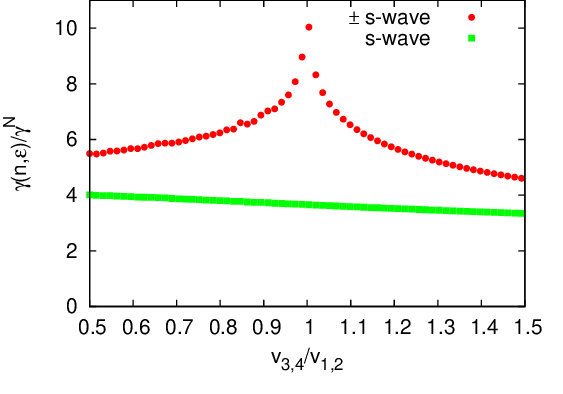}
\caption{\label{fig:fig10}
(Color online) Fermi velocity dependence of the normalized energy width $\tilde{\gamma}(\epsilon,n)$ with $n=4$ in the isotropic $\pm s$-wave and $s$-wave superconductors. The energy is $\epsilon = 0.1\Delta$.
}
\end{figure}
%

We discuss the dependence on the modulus of the Fermi velocity in the normalized energy width. 
As shown in Fig.~\ref{fig:fig10}, the energy width diverges when the amplitudes of the Fermi velocity on the each Fermi surface are same in the $\pm s$-wave superconductor. 
Therefore, the parameter $v_{3,4} = 0.99 v_{1,2}$ used in Sec.~V is in the range of application. 
 
\begin{figure}
  \begin{center}
    \begin{tabular}{p{50mm}p{30mm}}
      \resizebox{50mm}{!}{\includegraphics{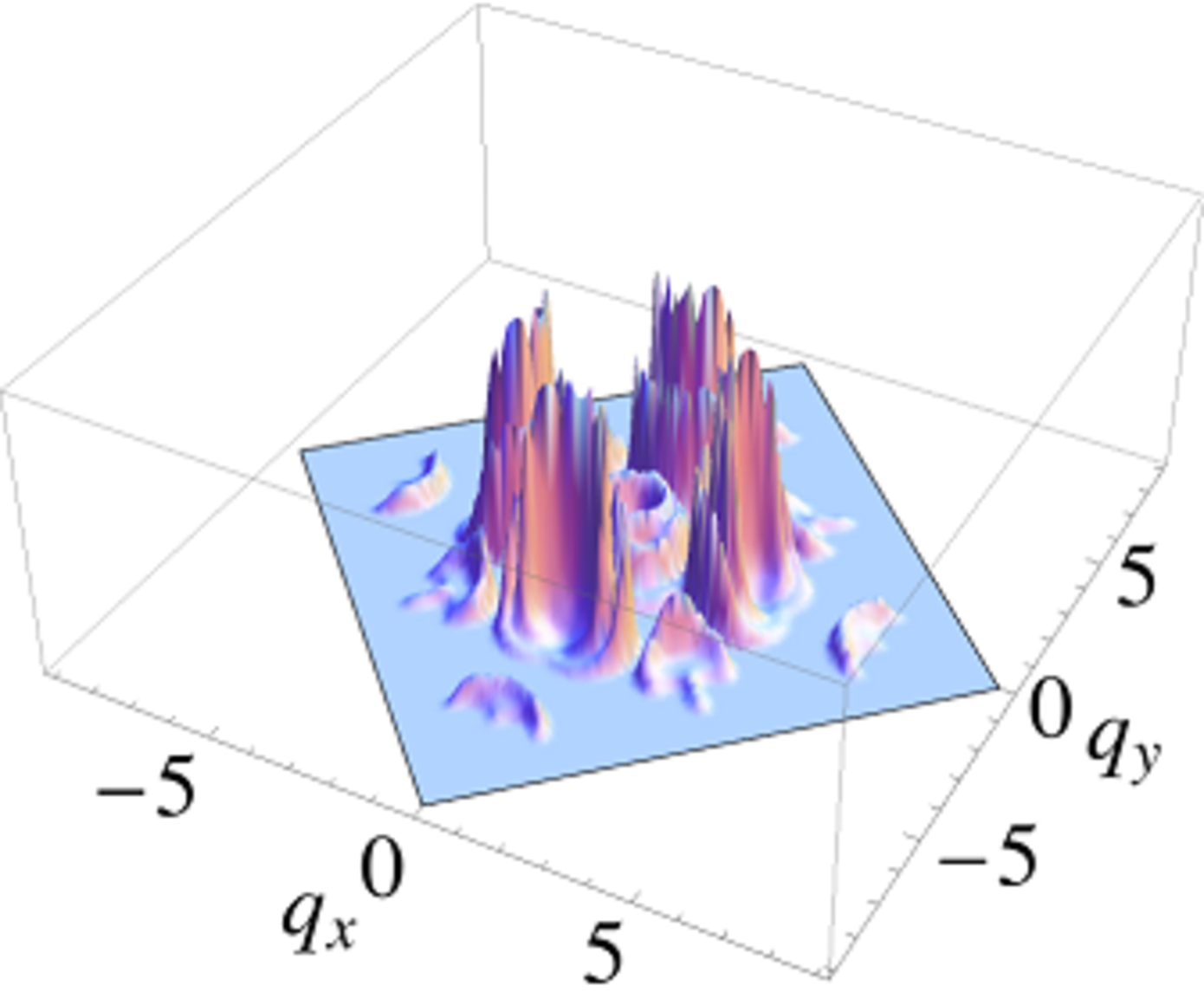}} &
      \resizebox{30mm}{!}{\includegraphics{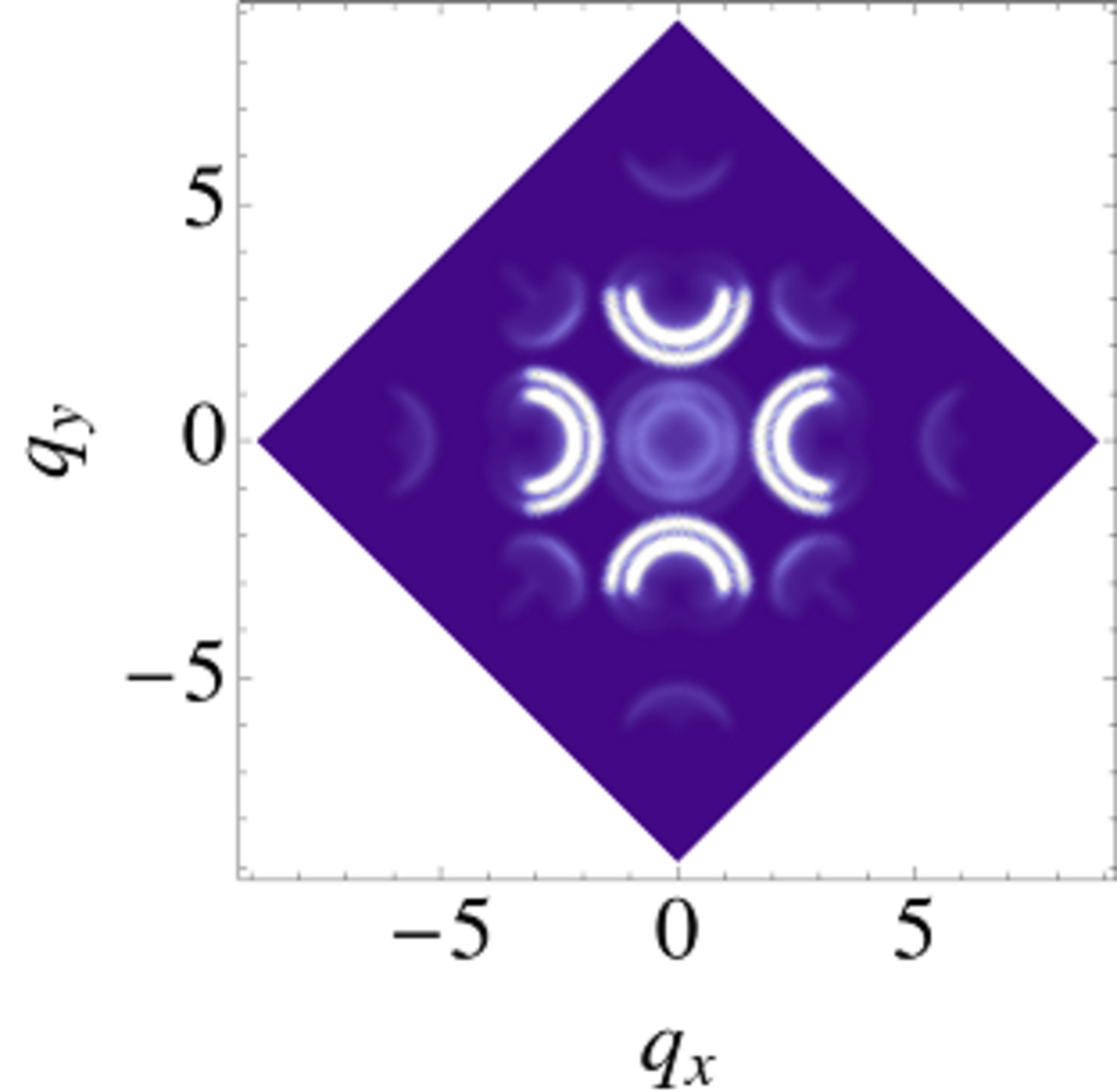}} 
    \end{tabular}
\caption{\label{fig:fig9}
(Color online)
$\Vec{q}$-dependence of the normalized impurity scattering rate $\tilde{\Gamma}_{\Vec{k},\Vec{k+q}}$ in the isotropic $\pm s$-wave superconductors. 
The energy is $\epsilon= 0.3 \Delta_{\infty}$. 
The amplitude of the Fermi velocity $v_{F,n}$ on $\alpha$ Fermi surface ($n = 1,2$) is $v_{F,n} = 0.8 v_{F,m}$ ($m = 3,4$).  
}
  \end{center}
\end{figure}

Overall momentum dependence in  $\tilde{\Gamma}_{\Vec{k},\Vec{k+q}}$ in the previous section do not change qualitatively as shown in Fig.~\ref{fig:fig9} even when the Fermi velocities are taken to be slightly different among the Fermi surfaces.
%

%

We have presented the results for two-dimensional systems because our study has been motivated by the Fe-based superconductors. 
However, there exist the Fe-based superconductors such as the 122- or 11-system \cite{Rotter,Hsu}, for which three-dimensionality of the band structure is important. We note that our formulation can be easily generalized to three-dimensional systems with use of the Fermi velocities projected on the plane perpendicular to a vortex.\cite{NagaiJPSJ} 
The Riccati equations (\ref{eq:a}) and (\ref{eq:b}) for three-dimensional systems turn into the same form as that for 
two-dimensional systems because of a translational symmetry along the direction of the vortex. 
Our results do not change qualitatively 
when the Fe-based superconductor has the three-dimensional Fermi surface, 
since the sign-change of a pair-potential in the momentum space is two-dimensional in the case of 
$\pm s$-wave and $d$-wave that we considered. 

We discuss the relevance of the characteristic impurity scattering rate in vortex cores in $\pm s$-wave superconductors to the QPI measurements.
For quasiparticles in the bulk, the Fourier-transform of the position-dependent conductance $g(\Vec{r},V)=dI(\Vec{r},V)/dV$ is proportional to the scattering rates or coherence factor of quasi-particles. If we assume this relation holds near vortex cores, then the impurity scattering rate for the Andreev bound state could be deduced from $g(\Vec{r},V)$ near vortex cores. Indeed, the spatial resolution of the STM/STS allows us to deduce the QPI patterns near vortex cores selectively\cite{Hanaguri}. 
If the very strong arc-like peaks were observed 
and the intensity near $\Vec{q} = 0$ is relatively small in the Fourier-transform of $g(\Vec{r},V)$, this would be a direct evidence for the $\pm s$-wave superconductivity. However, $g(\Vec{r},V)$ is suffering from extrinsic effects coming from the scanning feedback loop and hence the coherence factor has been deduced 
from the Fourier-transform of $Z(\Vec{r},V)=g(\Vec{r},V)/g(\Vec{r},-V)$ in Ref.~\onlinecite{Hanaguri}. 
Thus, for a direct comparison of our results with the experimental results on QPI, it is required to calculate the contribution from the Andreev bound states to $Z(\Vec{r},V)$ near vortex cores. A future problem is an explicit calculation of how novel properties of impurity scattering inside vortex cores of $\pm s$-wave superconductors are reflected in the QPI measurements. Our findings on the coherence effects inside vortex cores in the present paper would provide a physical interpretation of the experimental and theoretical results on $Z(\Vec{r},V)$.

\section{Conclusion}
We studied the impurity effects in vortex cores of various kinds of superconducting pairing symmetries. 
We found that the sign-reversed forward scatterings are dominant for these impurity scatterings by 
the Andreev bound states in the low energy. 
The $\pm s$-wave superconductivity 
yields %
strong arc-like peaks and weak intensity near $\Vec{q} = 0$ 
in the Fourier-transform of $dI/dV$ by the STM/STS measurements.  
We discussed the relevance of our results to the QPI measurements. 

\section*{Acknowledgment}
We thank T. Hanaguri and N. Hayashi for helpful discussions and comments. 
We also thank participants of the 12th International Workshop 
on Vortex Matter
in Superconductors at Lake Yamanaka in Japan on 2009 for various kinds of discussions and comments. 
Y.N.~acknowledges support 
   by Grant-in-Aid for JSPS Fellows (204840). 
This work is partially supported by the Ministry of Education, Science, Sports and Culture, Grant-in-Aid for Scientific Research on Priority Areas, 20029007, and also supported by Japan Society of Promotion of Science, Grant-in-Aid for Scientific Research (C), 21540352. 



\begin{thebibliography}{99}
\bibitem{note}
We introduce the self-energy and the Green function used in Ref.~\onlinecite{KopninText}
\end{thebibliography}
\end{document}